\documentclass[%
 table,
 superscriptaddress,
 amsmath,amssymb,
 aps,
 pra,
]{revtex4-2}

\usepackage[utf8]{inputenc}
\usepackage[T1]{fontenc}
\usepackage{amssymb}
\usepackage{mathtools}
\usepackage{bbold}
\usepackage{bm}
\usepackage[position=top, caption=false]{subfig}
\usepackage[x11names,dvipsnames,table,xcdraw]{xcolor}
\usepackage{tikz}
\usetikzlibrary {shapes.geometric}

\usepackage{amsthm}
\usepackage{amsmath}
\theoremstyle{definition}
\newtheorem{definition}{Definition}

\usepackage{array,hhline}

\begin{document}

\title{Adaptable Weighted Token Swapping Algorithm for Optimal Multi-Qubit Pathfinding}

\affiliation{School of Physics, University of Melbourne, VIC, Parkville, 3010, Australia.}

\author{Gary J Mooney}
\email{mooneyg@unimelb.edu.au}
\affiliation{School of Physics, University of Melbourne, VIC, Parkville, 3010, Australia.}

\date{\today}

\begin{abstract}
Quantum computing promises breakthroughs in simulating and solving complex, classically intractable problems. However, current noisy intermediate-scale quantum (NISQ) devices are relatively small and error-prone, prohibiting large-scale computations. To achieve quantum advantage in this regime, it is crucial to minimise the impact of noise from qubit decoherence and two-qubit gates. A direct approach is to optimise quantum circuit compilation, particularly by improving how circuits are mapped onto hardware. This work targets multi-qubit pathfinding (MQPF), a key subproblem in quantum circuit mapping, formulated as a variant of the token swapping problem. We propose an adaptable algorithm, modelled as a binary integer linear program, that routes $K$ teams of qubits on hardware graphs using swap operations. The algorithm minimises SWAP-gate depth and accumulated gate and idle errors, effectively solving a weighted version of the parallel~($K+1$)-coloured token swapping problem. We benchmark performance across various hardware layouts, comparing runtimes, SWAP depths, gate counts, and errors. Our results show that the proposed MQPF algorithm offers significantly improved runtime scaling and lower accumulated errors over a state-of-the-art exact SMT-CBS-based method. Potential applications include precomputing optimal routing for circuit mappers, benchmarking heuristics, and informing quantum hardware design by analysing pathfinding behaviour.
\end{abstract}

\maketitle

\section{Introduction} \label{sec:introduction}

Quantum computing technology is currently in the Noisy Intermediate-Scale Quantum (NISQ) era. Executing robust algorithms for realistic problem sizes remains challenging due to the significant noise in these devices, predominantly from two-qubit gate error rates and qubit decoherence. This noise limits the number of two-qubit gates that can be applied before the results are obfuscated. To combat these limitations and perform the largest algorithms possible on current quantum hardware, it is crucial to optimise quantum circuits during compilation to NISQ devices. This involves minimising both the circuit execution time and the number of two-qubit gates while prioritising low error rates.

There are two major compilation stages when mapping an arbitrary abstract quantum circuit to a target NISQ device. In the first stage, the original gates of the circuit are decomposed into base gates that are directly supported by the device. Typically, the set of possible base gates in a decomposed circuit consists of a two-qubit gate and parameterised single qubit gates. In the second stage of compilation, which we refer to as \textit{quantum circuit mapping}, the abstract circuit is mapped to the hardware graph of a NISQ device by representing each abstract qubit with a physical one. Implementing an abstract two-qubit gate on the device thus requires its corresponding physical qubits to be adjacent to one another. However, due to the limited connectivity of many NISQ-devices, abstract qubits in typical quantum circuits need to be frequently relocated to satisfy gate adjacency requirements. The standard approach for qubit relocation is to use sequences of SWAP gates, each constructed from three CNOT gates. Thus, mapping to the hardware graph can lead to a significant overhead in the number of CNOT gates and the CNOT depth of circuits. Optimising the quantum circuit during this stage can alleviate this overhead and significantly reduce the quantum resource requirements. 

\begin{figure}
     \centering
     \includegraphics[width=0.65\linewidth]{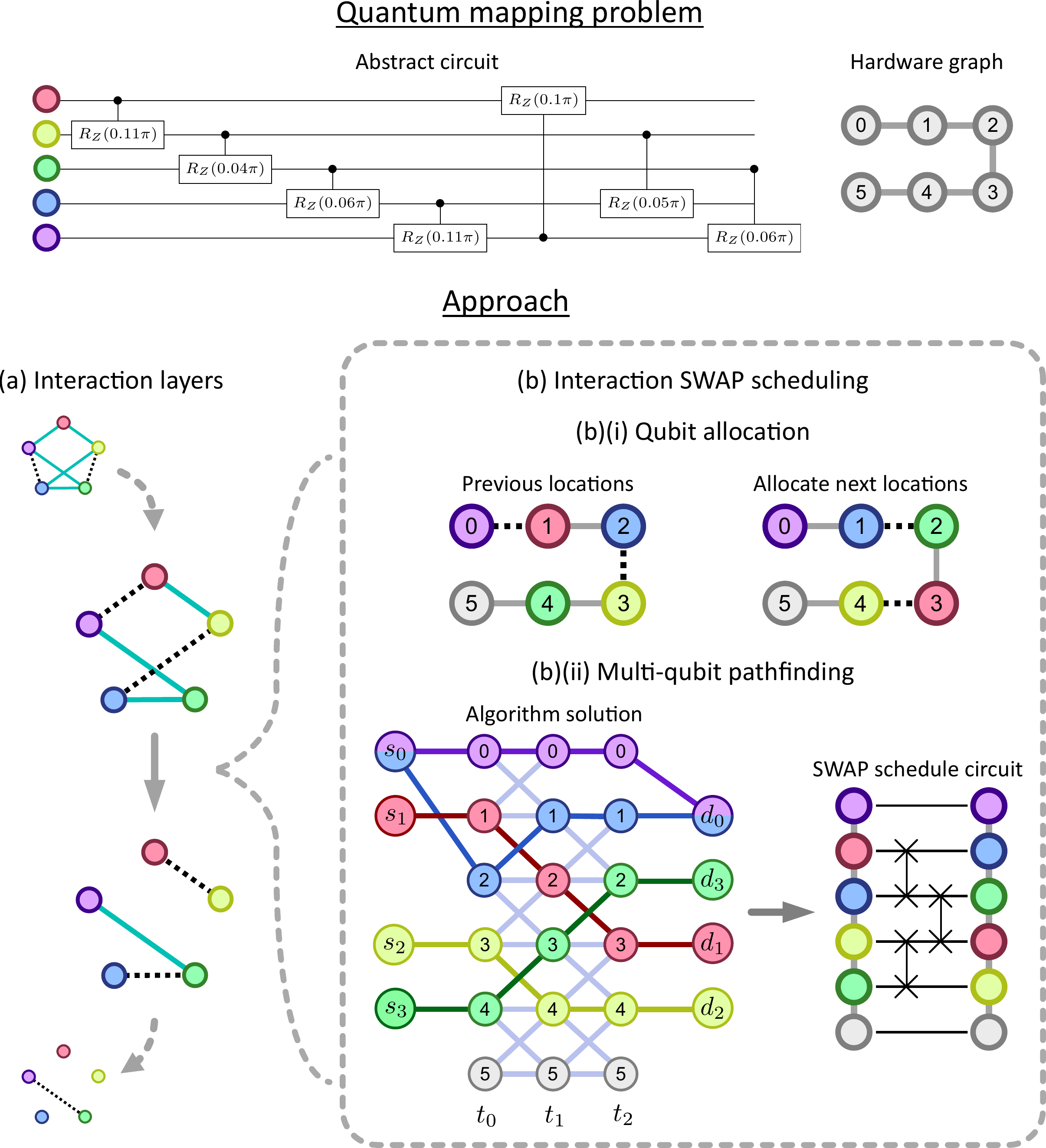}
    \caption{An overview of the approach used for the quantum mapping problem on an example problem instance. The abstract circuit is to be executed on the hardware that has limited connectivity. In order to perform the circuit's two-qubit gates directly on hardware, the involved qubits must be moved to be adjacent using sequences of SWAP gates. \textbf{(a)} An interaction graph represents the qubit-pair interactions that can be targeted next in the abstract circuit. This graph is updated in layers based on which pairs interact from the SWAP schedule and which are remaining in the abstract circuit. \textbf{(b)} The SWAP schedule process that routes qubits to interact at each layer. \textbf{(b)(i)} Qubit allocation is used to find the next set of destinations for the abstract qubits to move to. \textbf{(b)(ii)} Multi-qubit pathfinding is used to determine SWAP schedules that move qubits to their destinations.  \label{fig:approach-overview}}
\end{figure}

Quantum circuit mapping (see Fig.~\ref{fig:approach-overview}) involves mapping arbitrary abstract circuits to devices with restricted-connectivity qubit layouts via sequences of SWAP gates. The problem of determining optimal mappings has been shown to be NP-complete, both in cases of minimising the total number of SWAP gates and minimising the SWAP depth (makespan) of the final circuit~\cite{maslov2008quantum, botea2018complexity}. A variety of heuristic and optimal approaches to the quantum circuit mapping problem have been introduced in the quantum compilation literature~\cite{huang2023near, kusyk2021survey}. Widely adopted examples include the SABRE swap-based heuristic in Qiskit~\cite{li2019tackling, zou2024lightsabre}, Cambridge Quantum Computing's t$|\text{ket}\rangle$ compiler~\cite{cowtan2019qubit, sivarajah2020t}, and Rigetti Computing's Quil compiler~\cite{smith2020open}. Although many others exist, including algorithms based on integer linear programming (ILP)~\cite{bhattacharjee2017depth,bhattacharjee2019muqut}, search methods~\cite{zhang2021time, zulehner2018efficient,oddi2018greedy,tannu2019not}, boolean satisfiability~\cite{tan2020optimal, wille2019mapping}, pseudo-boolean optimisation~\cite{wille2014optimal, lye2015determining}, graph theory~\cite{nash2020quantum}, machine learning~\cite{li2023quantum}, recursive-subcircuit cutting~\cite{maslov2008quantum}, specific architecture tailoring~\cite{jin2021structured}, and temporal planning~\cite{booth2018comparing, venturelli2018compiling, venturelli2017temporal}, are all considered. As the size of quantum devices and circuits increases, optimal quantum circuit mapping quickly becomes intractable, even for short circuits on relatively small devices of around~10 qubits~\cite{tan2020optimal, wille2019mapping}. Consequently, in practice, optimality of the full quantum circuit mapping problem is often foregone in favour of applying combinations of heuristic techniques, approximations, or problem simplifications. A common simplification, shown in Fig.~\ref{fig:approach-overview}, involves splitting the circuit into interaction layers, where each layer contains the maximal number of abstract two-qubit interactions that can be processed in parallel~\cite{zulehner2018efficient, bhattacharjee2019muqut, cowtan2019qubit}. The challenge then reduces to finding a minimal SWAP gate schedule for each interaction layer to route qubits to facilitate the necessary two-qubit gates. This SWAP gate scheduling at each layer can be further simplified to two steps~\cite{maslov2008quantum, wille2019mapping,siraichi2019qubit,tan2020optimal}: qubit allocation and multi-qubit pathfinding. Qubit allocation~\cite{zhu2020exact, shafaei2014qubit,wille2019mapping}, involves assigning abstract qubits to physical qubits on the device hardware graph in order to satisfy adjacency requirements for the interaction layer. Multi-qubit pathfinding (MQPF), which is the focus of this paper, performs a permutation of abstract qubits on the hardware graph, routing them to their target locations using sequences of SWAP operations. Finding optimal routes with respect to quantum circuit resources and noise has a direct and significant impact on the performance of NISQ computations. Additionally, obtaining optimal MQPF SWAP sequences can be used to precompute optimal routing data that is utilised in quantum circuit mapping algorithms~\cite{wille2014optimal}, and could serve as a reference to benchmark non-optimal MQPF algorithms and NISQ architectures. To summarise this approach to full quantum circuit mapping, the problem is simplified by focusing on individual interaction layers in the abstract circuit and routing abstract qubits on the hardware graph using alternating applications of qubit allocation and MQPF. These subproblems are greatly reduced in size, leading to a considerable computational speedup overall compared to solving the quantum circuit mapping problem in full. 

\begin{figure}
     \centering
     \qquad
     \subfloat[][15q \textit{ibmq\_16\_melbourne}\label{fig:ibmq-melbourne}]{
         \centering
         \includegraphics[scale=1.4]{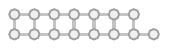}
     }
     \qquad\qquad
     \subfloat[][20q \textit{ibmq\_poughkeepsie}\label{fig:ibmq-poughkeepsie}]{
         \centering
         \qquad
         \includegraphics[scale=1.4]{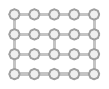}
        \qquad
     }
     \qquad
     \qquad
     \subfloat[][20q Rigetti \textit{Acorn}\label{fig:rigetti-acorn}]{
         \vspace{6pt}
         \centering
         \includegraphics[scale=1.4]{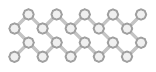}
     }
     \qquad
     \subfloat[][27q \textit{ibmq\_paris}\label{fig:ibmq-paris}]{
         \vspace{6pt}
         \centering
         \includegraphics[scale=1.4]{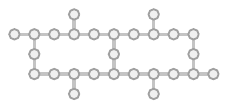}
     }
     \subfloat[][53q \textit{ibmq\_rochester}\label{fig:ibmq-rochester}]{
         \vspace{6pt}
         \centering
         \qquad
         \includegraphics[scale=1.4]{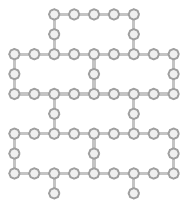}
         \qquad
     }
     \subfloat[][64q 8x8 grid\label{fig:grid-8x8}]{
         \vspace{6pt}
         \centering
         \includegraphics[scale=1.4]{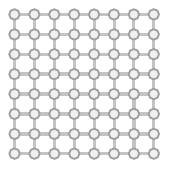}
     }
        \caption[The target hardware graphs used for benchmarking the multi-qubit pathfinding algorithms in this work.]{The target hardware graphs used for benchmarking the multi-qubit pathfinding algorithms in this work. \textbf{(a)} The 15-qubit \textit{ibmq\_16\_melbourne} device~\cite{ibmq}. \textbf{(b)} The 20-qubit \textit{ibmq\_poughkeepsie} device. \textbf{(c)} The 20-qubit Rigetti \textit{Acorn} device~\cite{otterbach2017unsupervised}. In the actual device, one of the qubits is offline, resulting in 19 working qubits. \textbf{(d)} The 27-qubit \textit{ibmq\_paris} device. \textbf{(e)} The 53-qubit \textit{ibmq\_rochester} device. \textbf{(f)} A 64-qubit 8x8 Grid.\label{fig:hardware-graphs}}
\end{figure}

Targeting the MQPF subproblem, we present a SWAP-depth optimal MQPF algorithm that secondarily minimises accumulated gate and idle error while achieving practical runtimes on current quantum hardware. The algorithm is modelled using binary integer linear programming (BILP) and initially optimises for the SWAP depth of the circuit as a primary objective, and then optimises for the accumulated error rate as a secondary objective. The algorithm supports multiple teams of qubits, each sharing its own set of destinations, where qubits in a team are indifferent to which of the team's assigned destinations they reach. From the perspective of the token swapping problem, for $K$ teams of qubits, the algorithm solves a weighted variant of the parallel ($K$+1)-coloured token swapping problem, which we discuss in Section~\ref{sec:token-swapping-problem}. However, our algorithm can additionally support the number of destinations for a team being greater than the number of qubits in the team, allowing the qubits to be routed to any subset of destinations. This also enables multiple teams to share the same destination in their destination sets at the same time. Properties of the algorithm's outputs, such as computational runtimes, SWAP depths, SWAP gate counts and accumulated error rates, are compared and analysed across the variety of device qubit layouts shown in Figure~\ref{fig:hardware-graphs}. We also compare three levels of optimiser settings of our MQPF algorithm with a state-of-the-art exact token swapping algorithm based on satisfiability modulo theories and conflict-based search (SMT-CBS)~\cite{surynek2019multi, surynek2019conflict}. For these comparisons, the error model is extended by including idle errors and assigning error rates to match the current IBM Quantum Heron devices. These results demonstrate significantly improved runtime scaling (3600 sec down to 1 sec in some larger instances) and solution SWAP schedule fidelities on random problem instances. Although the error maps and hardware layouts targeted in this work are based on superconducting qubit technology, this approach is optimal for any platform that primarily uses SWAP gates to move abstract qubits in order to satisfy connectivity constraints, where the SWAP-gate and idle-time errors are large enough that we can assume they are the only sources of error we need to consider in the schedule. The algorithm can be adapted to support SWAP-gate limitations at individual time steps if the hardware requires it.

In Section~\ref{sec:mqp-preliminaries}, we discuss relevant existing techniques for modelling variations of multi-agent pathfinding problems. In Section~\ref{sec:mqp-multi-qubit-pathfinding}, we formally define the MQPF problem and discuss the steps involved in modelling it using a BILP approach. In Section~\ref{sec:results}, we describe the details of the two experimental implementations and present the results. In Section~\ref{sec:mqp-discussion}, we conclude with a discussion of the results.

\section{Techniques Toward Multi-Qubit Pathfinding}\label{sec:mqp-preliminaries}

\subsection{Linear Programming}
Linear programming is a widely used optimisation method for problems that can be represented as a set of linear constraints with real variables, aimed at maximising or minimising a specified linear cost function, also commonly referred to as the objective function. LP is extensively used to model a broad array of real-world problems, including various scheduling, resource allocation, transportation, and routing problems. The power of this method comes from its capability to find optimal solutions in polynomial time, combined with the flexibility it offers in application---especially when simplifying assumptions are integrated into the model. 

The standard form of an LP can be written as
\begin{align}
    &\max {\boldsymbol{c} \cdot \boldsymbol{x}},\\
    \text{such that } &A\boldsymbol{x} \leq \boldsymbol{b},\\
    \text{and }&x_i \geq 0,\\
    \text{where }&x_i \in \mathbb{R},\;\; i=0, 1, \ldots n,
\end{align}
where $\boldsymbol{c} = (c_1,c_2, \ldots, c_n)\in \mathbb{R}^n$ is the coefficient vector of the objective function, $\boldsymbol{x} = (x_1,x_2,\ldots,x_n)\in \mathbb{R}^n$ is the vector of variables to be determined, and $\boldsymbol{b}=(b_1, b_2 \ldots, b_m)\in \mathbb{R}^m$ is the vector of constants defining the upper bounds in the constraints. The constraint matrix $A$ is given by: 
\begin{equation}
    A = \left(\begin{array}{ccc}
        a_{1,1} & \ldots & a_{1,n} \\
        \vdots & \ddots & \vdots \\
        a_{m,1} & \ldots & a_{m,n}
    \end{array}\right),
\end{equation}
where each element $a_{i,j}\in\mathbb{R}$ specifies the impact of variable $x_j$ on the $i$-th constraint.

This formulation allows LP to be a powerful tool, as it provides a structured way to optimise complex systems under constraints, ensuring that all specified conditions are met while achieving the best possible outcome with respect to the objective function.

\subsection{Binary Integer Linear Programming (BILP)}
In many real-world applications, the variables we aim to optimise over are not continuous but must be restricted to binary or integer values. This occurs when optimising with respect to discrete quantities, such as the number of cars produced, or decisions that involve logical relations, like whether to proceed with Project A if it precludes Projects B and C. For such cases, extended forms of mathematical programming come into play such as pure integer, binary integer, and mixed integer linear programming. Pure integer linear programs constrain all variables to integer values, binary integer linear programs (BILPs) restrict them to binary values, and mixed integer linear programs include both integer and real variables. Generally, linear programming problems that include variables that are limited to integer values (including binary) belong to the NP-hard complexity class~\cite{kannan1978computational}. This work focuses on BILP, which can be applied to a wide range of problems including the best choice problem (or secretary problem)~\cite{ferguson1989solved}, the travelling salesman problem, satisfiability problems, generalised assignment problems, and minimum weight cut problems~\cite{conforti2014integer, vanderbei2015linear}.

The standard form for a BILP can be written as 
\begin{align}
    &\max {\boldsymbol{c} \cdot \boldsymbol{x}},\\
    \text{such that } &A\boldsymbol{x} \leq \boldsymbol{b},\\
    \text{where }&x_i \in \{0,1\},\;\; i=0, 1, \ldots n,
\end{align}
where $\boldsymbol{c} = (c_1,c_2, \ldots, c_n)\in \mathbb{R}^n$ are the coefficients of the objective function, $\boldsymbol{x} = (x_1,x_2,\ldots,x_n)\in \{0,1\}^n$ denotes the binary variables under consideration, and $\boldsymbol{b}=(b_1,b_2, \ldots, b_m)\in \mathbb{R}^m$ is the vector of constants defining the upper bounds in the constraints. The constraint matrix $A$ is given by: 
\begin{equation}
    A = \left(\begin{array}{ccc}
        a_{1,1} & \ldots & a_{1,n} \\
        \vdots & \ddots & \vdots \\
        a_{m,1} & \ldots & a_{m,n}
    \end{array}\right),
\end{equation}
where each element $a_{i,j}\in\mathbb{R}$ specifies the impact of variable $x_j$ on the $i$-th constraint. While finding exact solutions to BILP problems is NP-hard, there exist efficient algorithms capable of approximating solutions with reasonable computational effort and accuracy~\cite{li2020simple, johnson2000progress, wedelin1995algorithm}.

\subsection{Multi-Agent Pathfinding (MAPF)}\label{sec:multi-qubit-pathfinding}

Multi-agent pathfinding (MAPF) is relevant to various disciplines and optimisation problems. While single-agent scenarios can utilise fast and efficient algorithms such as the $A^*$ pathfinding algorithm~\cite{hart1968formal} or the Dijkstra's pathfinding algorithm, the introduction of multiple agents introduces complexities due to the need for cooperation to avoid collisions and minimise interference with each other's trajectories to achieve a globally optimal solution. Here, we focus on a problem referred to as time-dependent discrete routing on a collision-free unit-distance graph. It is characterised by discrete time steps where agents can complete any of their actions in exactly one time step, only one agent can occupy a single node at any time step, and agents cannot pass through each other.

Following the work in~\cite{ma2016optimal}, the MAPF problem can be split into two primary variants: anonymous and non-anonymous pathfinding. Ideas from these two extremes can be generalised to form the combined target-assignment and pathfinding (TAPF) problem.

\subsubsection{Anonymous Multi-Agent Pathfinding}

\begin{figure}
     \centering
     \subfloat[][Hardware graph\label{fig:multi-agent-pathfinding-initial-graph}]{
         \centering
         \includegraphics[scale=1.2]{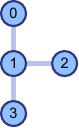}
         \qquad\qquad
     }
     \qquad
     \subfloat[][Edge transformation\label{fig:multi-agent-pathfinding-movement-transformation}]{
         \centering
         \includegraphics[scale=1.2]{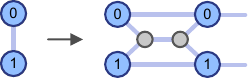}
     }
     \qquad
     \subfloat[][Time-expanded graph\label{fig:multi-agent-pathfinding-time-expanded-graph}]{
         \vspace{6pt}
         \centering
         \includegraphics[scale=1.2]{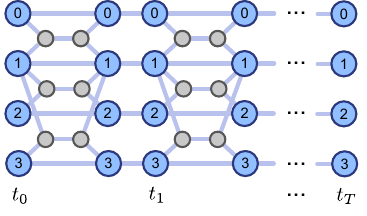}
     }
        \caption[An example of the anonymous multi-agent pathfinding time expansion graph transformation.]{An example of the anonymous multi-agent pathfinding time expansion graph transformation. \textbf{(a)} An undirected simple input graph $G$ describing the possible agent locations and movements corresponding to the hardware for a particular problem instance. \textbf{(b)} To expand the graph through time, an edge in the original graph can be transformed as shown. This transformation enables agents to either be idle or move to an adjacent node for the next time step. The additional two grey nodes enforce the rule that only one agent can use the movement edge per time step, this is required to stop the agents moving through each other. The outgoing edges to the very right are used to enforce that only one agent can be at any one location per time step. \textbf{(c)} The input graph is converted to a time-expanded graph with $T$ time steps labelled $t_0, t_1, \ldots t_T$. Two copies of the original input graph nodes are used for each time step layer, along with an additional two nodes for each edge. \label{fig:multi-agent-pathfinding-time-expand}}
\end{figure}

In the anonymous variant of MAPF, agents are indistinguishable from one another and are collectively assigned to a set of destinations without specific assignments per agent. Consider a connected undirected graph $G(V, E)$ with nodes $V$ representing possible locations and edges $E$ representing possible movements, and each node includes a loop at allow agents to remain idle. Given $N$ agents $\boldsymbol{A}$, with $N$ source nodes $\boldsymbol{S}$ and $N$ destination nodes $\boldsymbol{D}$, the objective is to find optimal paths $p_a:\mathbb{Z}\rightarrow V$ for each agent $a\in \boldsymbol{A}$ that satisfy a range of conditions. The paths must map time step indices $0, 1, \ldots, T$ to nodes such that consecutive positions $p_a(t)$ and $p_a(t+1)$ are either adjacent in $G$ or identical (implying the agent remains idle). Additionally, the paths can revisit nodes, thus forming \textit{trails} in graph-theoretical terms. The paths for the anonymous multi-agent pathfinding problem must satisfy the following conditions:

\begin{itemize}
    \item \textbf{Exclusivity of location:} No two agents can occupy the same location at the same time step.\\ $\forall a,b \in \boldsymbol{A}$ \text{where }$a\neq b,$ \text{and }$ \forall t \in \{0, 1, \ldots T\}$, \text{then }$p_a(t) \neq p_b(t)$.
    \item \textbf{Non-intersecting paths:} Agents cannot pass through each other.\\
    $\forall a,b \in \boldsymbol{A}$ \text{where }$a\neq b,$ \text{and }$ \forall t \in \{0, 1, \ldots T\}$, \text{then} $\big(p_a(t), p_b(t)\big) \neq \big(p_b(t+1), p_a(t+1)\big)$.
    \item \textbf{Starting conditions:} Each agent starts at its assigned source location.\\
    $\forall a \in \boldsymbol{A}$, $\exists s\in\boldsymbol{S}$, \text{such that} $p_a(0)=s$.
    \item \textbf{Ending conditions:} Each agent ends at a destination location.\\
    $\forall a \in \boldsymbol{A}$, $\exists d\in\boldsymbol{D}$, \text{such that} $p_a(T)=d$.
\end{itemize}
The goal is to find valid paths $p_a$ for each agent $a$ that minimises the makespan $T$ (number of time steps). This form of pathfinding can be solved using discrete dynamic network flow techniques~\cite{yu2013multi}, transforming the input graph $G$ into a time-expanded graph as shown in Figure~\ref{fig:multi-agent-pathfinding-time-expand}. Time steps are layered left to right in order of increasing time. Each time step consists of two copies of nodes $V$ of $G$ along with two additional nodes for each possible agent movement, as specified by edges in $G$. The edges in the time-expanded graph are determined from the transformation shown in Figure~\ref{fig:multi-agent-pathfinding-time-expand}b. These edges facilitate rules preventing agents from moving through each other and ensure that only one agent occupies a location at any given time step, while permitting agents to either remain idle or move to an adjacent node. This problem can be modelled as a max flow or min-cost flow problem on the time-expanded graph by specifying source nodes in the initial layer $t=0$ for starting locations and sink nodes in the final layer $t=T$ for end locations, with the edge capacities set to 1 to restrict each edge in the time-expanded graph to a single agent, and adding traversal costs as edge weights if using min-code flow to minimise total traversal costs. The corresponding max flow or min-cost flow algorithms, such as the fast and efficient Edmonds-Karp algorithm~\cite{edmonds1972theoretical}, are performed incrementally on increasing values of maximum time step $T$, starting at $T=0$, until a valid solution is found, indicating the optimal makespan. It has been shown that there will always be a feasible solution (as long as the graph $G$ is connected) and that the optimal makespan is always no larger than $N + |V| - 1$, where $|V|$ is the number of nodes~\cite{yu2013multi}.

\subsubsection{Non-Anonymous Multi-Agent Pathfinding}

In the non-anonymous variant of MAPF, each agent is uniquely identified and assigned to a specific target destination, which is not shared with any other agent. This problem is prevalent in fields such as video games, robotics, and automated warehouse systems~\cite{wang2008fast,kavraki1996probabilistic,luna2010network,basile2012hybrid}. The problem is NP-hard and the optimal makespan solution is bounded by $\mathcal{O}(|V|^3)$. There are a variety of existing strategies for solving this problem, including ILP~\cite{yu2013planning}, satisfiability~\cite{surynek2015reduced}, answer set programming~\cite{erdem2013general}, and search based methods~\cite{standley2010finding, sharon2013increasing, cohen2015feasibility, sharon2015conflict, boyarski2015icbs}.

\subsubsection{Combined Target-Assignment and Pathfinding (TAPF)}

The ideas of anonymous and non-anonymous MAPF can be generalised with each representing extreme cases of the broader problem. The TAPF problem involves assigning target destinations to agents and subsequently finding collision-free paths to these destinations, all while minimising the makespan. There are multiple teams of agents, where each team is allocated a set of destinations that can be shared internally but not with agents from other teams. As a generalisation of non-anonymous MAPF, TAPF is also NP-hard (even for two teams) and the optimal makespan solution remains bounded by $\mathcal{O}(|V|^3)$. There are various existing approaches to solve TAPF, including conflict-based min-cost flow (CBM)~\cite{ma2016optimal}, multi-commodity flow~\cite{ma2016optimal,yu2013planning}, and answer set programming~\cite{nguyen2019generalized}. 

The CBM approach is a particularly popular approach to TAPF. It is a hierarchical algorithm with distinct high and low levels of processing. At the higher level it considers each team as a single meta-agent and uses conflict-based search (CBS)~\cite{sharon2015conflict, honig2018conflict} to resolve collisions between each team. CBS uses a breadth-first search on a tree where each node in the tree contains paths for each team alongside a set of constraints. When collisions occur, the tree is expanded with nodes that apply direct constraints to avoid the conflict---for example, if team 1 and 2 have a collision at an edge, two nodes are added: one restricts team 1 from using the edge, and the other restricts team 2. For each node, new paths are recalculated for the teams with their additional constraints. At the lower level, paths for each team are determined using an anonymous multi-agent pathfinding algorithm, which can be performed in polynomial time. 

Another approach, of which is particularly interesting to the application of qubits, is the multi-commodity flow method, which addresses scenarios of max flow or min-cost flow where there is more than one commodity. Each commodity has its own flow demand with different source and destination nodes, and the flow capacity of network edges is shared among all commodities. In the following sections, we demonstrate how multi-commodity flow can be adapted into a BILP to effectively model the \textit{multi-qubit pathfinding problem}.

\subsection{The Token Swapping Problem (TSP)}\label{sec:token-swapping-problem}

In physical quantum devices, the movement of abstract qubits---each encoded as the state of a single physical qubit---is primarily actioned using SWAP gates. Consequently, in the context of quantum computing, pathfinding typically involves swap-based movement of abstract qubits. The makespan of the resulting solution directly corresponds to the SWAP depth of the mapped circuit. Within the computer science community, this problem is a variant of the more generally known token swapping problem (TSP). 

The TSP, as defined in the literature~\cite{alon1994routing, yamanaka2015swapping}, involves placing a token on each node of a connected simple graph $G$. The objective is to manoeuvre these tokens from their initial positions to assigned destinations via swap operations, minimising the total number of swaps. Only one pair of tokens can be swapped at a time. The TSP is generally NP-hard~\cite{kawahara2017time, miltzow2016approximation, bonnet2018complexity} but has been solved in polynomial time on specific graph structures such as paths, cycles, complete graphs, and complete bipartite graphs~\cite{yamanaka2015swapping, jerrum1985complexity, cayley1849lxxvii}.

There are many variants to the TSP. A notable variant is the $c$\textit{-coloured token swapping problem} ($c$-CTSP)~\cite{yamanaka2015swapping}, where tokens and nodes are assigned one of $c$ possible colours. The goal here is to manoeuvre the tokens such that each node has a token of its assigned colour, essentially routing coloured tokens to achieve a permutation to the final node-colour configuration starting from the initial token-colour configuration. The 2-CTSP has been shown to be solvable in polynomial time, whereas the 3-CTSP is NP-complete~\cite{yamanaka2015swapping}.  Another variant is the \textit{parallel token swapping problem} (PTSP)~\cite{kawahara2017time}, which permits any number of disjoint pairs of tokens to be simultaneously swapped in the same time step, unlike the TSP that only allows a single swap operation at a time. Like the TSP, the PTSP has been shown to be NP-complete. These two variants can be combined to form the \textit{parallel $c$-coloured token swapping problem} ($c$-PCTSP), which is NP-hard, even when limited to 2-PCTSP with only two colours. 

\section{Multi-Qubit Pathfinding (MQPF)}\label{sec:mqp-multi-qubit-pathfinding}
In this section, we detail the MQPF problem and introduce our MQPF algorithm that solves it. In quantum compilation literature, the MQPF problem usually refers to certain variations of the token swapping problem~(TSP). The primary MQPF problem addressed in our work for $K$ teams of qubits can be summarised as the weighted variant of the parallel ($K+1$)-coloured token swapping problem ($(K+1)$-PCTSP), discussed in the previous section. The extra colour represents physical qubit locations not occupied by abstract qubit states. This problem is NP-hard, even for a single team.

Current algorithms for solving MQPF exactly are based on recursive subcircuit cutting~\cite{maslov2008quantum}, satisfiability modulo theories and conflict-based search~\cite{surynek2019multi, surynek2019conflict}, search-based methods~\cite{miltzow2016approximation}, Pseudo-Boolean Optimisation~\cite{lye2015determining} and ILP (independently developed to our work, using a different modelling approach)~\cite{bhattacharjee2017depth}, and approximation methods that minimise noise~\cite{sharma2023noise}. These techniques typically aim to optimise either the SWAP gate count or the SWAP depth---total number of time steps---in the mapped routing circuit, and with each abstract qubit operating as its own independent team.

Given the prevalence of noise introduced through qubit decoherence and the application of two-qubit gates on physical quantum devices, it is important to minimise the computation time and the total error introduced. To do this, our model focuses on minimising the SWAP depth primarily, followed by minimising the accumulated error as a secondary objective. In addition to solving this problem, our algorithm supports flexibility in the assignment of abstract qubit teams and destination nodes. Teams of qubits can share destinations, and the number of destinations for each team can exceed the number of qubits in the team. The teams can then be routed to any subset of their assigned destinations.

We formally define the MQPF problem as follows:
\begin{definition}[The MQPF Problem]
Consider a connected hardware graph $G(V,E)$, where nodes $V$ correspond to qubits and edges $E$ correspond to possible applications of two-qubit gates (such as CNOTs and SWAPs). Given $N$ abstract qubits, partitioned into $K$ disjoint teams $\boldsymbol{A} = \boldsymbol{A}_1 \cup \boldsymbol{A}_2 \cup\ldots \boldsymbol{A}_K$, with each team assigned disjoint sets of source and destination locations $\boldsymbol{S}=\boldsymbol{S}_1 \cup \boldsymbol{S}_2 \cup\ldots \boldsymbol{S}_K$ and $\boldsymbol{D} = \boldsymbol{D}_1 \cup \boldsymbol{D}_2 \cup\ldots \boldsymbol{D}_K$, where $|\boldsymbol{A}_k| = |\boldsymbol{S}_k| = |\boldsymbol{D}_k|$ (although the model allows for $|\boldsymbol{A}_k| \leq |\boldsymbol{D}_k|$ and overlapping destination partitions $\boldsymbol{D}_k$). For each abstract qubit $a \in \boldsymbol{A}$, find a path through time $p_a:\mathbb{Z}\rightarrow V$ which maps time step indices $0, 1, \ldots T$ to nodes in $G$ such that for all time steps $t$, either $p_a(t)$ and $p_a(t+1)$ are adjacent in $G$ or $p_a(t) = p_a(t+1)$. Each path must satisfy the following conditions:
\begin{itemize}
    \item \textbf{Exclusivity of location:} No two abstract qubits can occupy the same location at the same time step.\\ $\forall a,b \in \boldsymbol{A}$ such that $a\neq b,$ and $ \forall t \in \{0, 1, \ldots T\}$, then $p_a(t) \neq p_b(t)$.\\
    \item \textbf{Swap-based movement:} Abstract qubits must swap when moving to an occupied location.\\
    $\forall a\in \boldsymbol{A}$ and $ \forall t \in \{0, 1, \ldots T\}$, if $\exists b \in \boldsymbol{A}$ such that $p_a(t+1) = p_b(t)$, then $p_b(t+1) = p_a(t)$.
    \item \textbf{Starting conditions:} All abstract qubits must start at sources corresponding to their team.\\
    $\forall a_k \in \boldsymbol{A}_k$, $\exists s_k\in\boldsymbol{S}_k$, such that $p_{a_k}(0)=s_k$.
    \item \textbf{Ending conditions:} All abstract qubits must end at destinations corresponding to their team.\\
    $\forall a_k \in \boldsymbol{A}_k$, $\exists d_k\in\boldsymbol{D}_k$, such that $p_{a_k}(T)=d_k$.
\end{itemize}
The primary objective of the problem is to find valid paths through time $p_a$ for each abstract qubit $a$ that minimises the SWAP depth, $T$ (number of time steps). For the secondary objective, error rates are assigned to each possible abstract qubit action in the hardware graph, aiming to minimise the combined total error rate of abstract qubit paths,~$\mathcal{E}$.
\end{definition}

\subsection{The MQPF Algorithm}\label{sec:mqpf_algorithm}

\begin{figure}
     \centering
     \subfloat[][Hardware graph\label{fig:time-expand-example-initial-graph}]{
         \centering
         \qquad\qquad\qquad
         \includegraphics[scale=1.2]{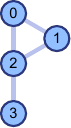}
         \qquad\qquad\qquad
     }
     \qquad
     \subfloat[][Time-expanded graph\label{fig:time-expand-example-expansion}]{
         \centering
         \includegraphics[scale=1.2]{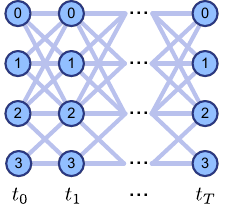}
     }
     \qquad
     \subfloat[][Attached sources and destinations\label{fig:time-expand-example-sources-and-destinations}]{
         \vspace{6pt}
         \centering
         \includegraphics[scale=1.2]{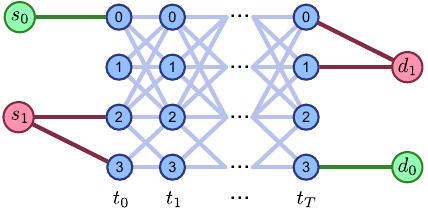}
     }
     \subfloat[][Solution paths\label{fig:time-expand-example-solution}]{
         \vspace{6pt}
         \centering
         \includegraphics[scale=1.2]{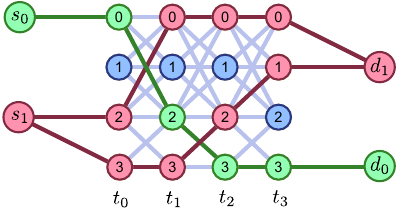}
     }
        \caption[An example of the process used when modelling a problem instance.]{An example of the process used when modelling a problem instance. \textbf{(a)} The initial hardware layout where nodes represent physical qubits and edges represent connected pairs of qubits that could be operated on by entanglement generating gates such as CNOTs. \textbf{(b)} The initial layout is converted to a time-expanded graph where layers represent time steps (where a time step is the amount of time required to perform a SWAP operation) labelled $t_0$, $t_1$, $\ldots$ $t_T$ and edges represent possible movements that abstract qubits could take between them. The possible movements only include being idle and swapping locations between connected physical qubits in the hardware layout. \textbf{(c)} Source and destination locations can be specified in the time-expanded graph by attaching special nodes ($s_0$, $s_1$, $d_0$, $d_1$) to the initial and final layers respectively. In this example, there are two teams of abstract qubits to be routed. Team~0 contains one abstract qubit starting at q0 and ending at q3 while Team~1 contains two abstract qubits starting at q2 and q3, and ending at q0 and q1. \textbf{(d)} The solution to this example problem instance with a SWAP depth of three (since the route has a total of three time steps).\label{fig:time-expand-example}}
\end{figure}

The general strategy of our algorithm is highlighted in Figure~\ref{fig:time-expand-example}. We begin by time expanding the input hardware graph as shown in Figure~\ref{fig:time-expand-example}b, similar to the approach used for anonymous multi-agent pathfinding discussed in Section~\ref{sec:multi-qubit-pathfinding}. In this expansion, each vertical layer, progressing from left to right, duplicates the hardware graph nodes and represents a single time step---defined as the duration required to execute a SWAP gate. These layers are labelled $t_0, t_1, \ldots t_T$. Each edge in the expansion represents possible movements that an abstract qubit can undertake when progressing through time steps; this includes remaining idle or swapping locations with adjacent physical qubits in the hardware graph. 

The time expansion process for MQPF is more straightforward than for MAPF because we model it using binary integer linear programming (BILP). This approach allows the constraints to be programmed directly into the model rather than using gadgets in the expanded graph. Once the hardware graph has been time expanded, we define teams of source and destination locations at the initial and final layers of the expanded graph, respectively. To accomplish this, special source $\boldsymbol{I}=\{I_1, I_2,\ldots I_K\}$ and destination $\boldsymbol{F}=\{F_1, F_2,\ldots F_K\}$ nodes, representing teams, are attached to the initial and final layers. These nodes are linked to their respective team source and destination locations, as demonstrated in Figure~\ref{fig:time-expand-example}c. 

To determine the optimal paths for abstract qubits through time, corresponding to the minimal SWAP depth, the algorithm initially expands to a depth $T=0$ and uses a BILP solver to search for a solution. If no solution is found, the graph is expanded to $T=1$, and the process is repeated. This iterative expansion and solution search continues until a solution is discovered, at which point the depth $T$ is confirmed as the optimal SWAP depth. It has been shown that a solution will always exist and the number of time steps is bounded by $\mathcal{O} ({|V|^2})$, where $|V|$ is the number of nodes in the graph $G$~\cite{yamanaka2016computational}.

At each time step, we aim to minimise the accumulated error $\mathcal{E}$ when solving the BILP. The accumulated error rate of the SWAP gate schedule is multiplicative, calculated as
\begin{equation}
 \mathcal{E} := 1-\prod_{\alpha\in \{\text{Movements}\}}(1-\epsilon_\alpha),
\end{equation}
where $\epsilon_\alpha$ is the error rate of the swap or idle movement $\alpha$ of an abstract qubit in the time expanded graph. However,~$\mathcal{E}$ is a product of variables while the BILP objective function must be linear. We can transform $\mathcal{E}$ into an equivalent linear objective function by taking $C :=-\log(1-\mathcal{E})$. This does not affect the solution because minimising $C$ is equivalent to minimising $\mathcal{E}$ due to the monotonicity of the $\log$ function. When substituting the expression for the accumulated error $\mathcal{E}$ into the objective function $C$, the $1-\mathcal{E}$ cancels the additive 1 value in $\mathcal{E}$ becoming only a product of values. This enables the objective $C$ to simplify to a sum of $\log$s
\begin{align}
    C &= -\sum_{\alpha\in \{\text{Movements}\}} \log(1-\epsilon_\alpha),\label{eq:objective_value}\\
    &= \sum_{\alpha\in \{\text{Movements}\}} c_\alpha
\end{align}
where $c_\alpha$ are the corresponding costs for each swap or idle movement $\alpha$ used in the solution. The objective value can be converted back to the accumulated error rate of the SWAP gate schedule using
\begin{equation}
    \mathcal{E} = 1-1/\exp(C).
\end{equation}

In Section~\ref{sec:mqp-results} we use a simple error model that includes only SWAP gate errors, $\mathcal{E}_{\text{SWAP}} = 1-\prod_{\alpha\in \{\text{SWAPs}\}}(1-\epsilon^\text{CNOT}_\alpha)^3$, where $\epsilon^\text{CNOT}_\alpha$ is the error rate of a CNOT used to perform the SWAP gate~$\alpha$ and $(1-\epsilon^\text{CNOT}_\alpha)^3$ is the success rate of the SWAP gate (since three CNOTs are used for each SWAP).

In Section~\ref{sec:comparison_with_smtcbs}, the error model is extended to additionally include idle errors by adding a cost to idle movement edges in the time expanded graph. Note that single-qubit gate errors can be ignored since there are no single-qubit gates in SWAP schedules, and measurement errors can be ignored since destination nodes are typically fixed for each problem instance (although they could be included by adding a constant to the objective function and the value will depend on the application of measurement error mitigation). The idle errors $\epsilon_i^\text{idle}$ are the errors introduced by the idle time of a CNOT gate for each idle edge $i$ used by an abstract qubit in the time expanded graph. The idle errors are calculated in the same way as done in mapomatic~\cite{PRXQuantum.4.010327}---a Python package used to find low noise subgraphs using the VF2 mapper algorithm~\cite{cordella2004sub}---which uses the idle time and the relaxation $T^{(i)}_1$ and dephasing $T^{(i)}_2$ decoherence times of the physical qubit $i$. The idle error for each idle qubit $i$ in a time step is evaluated as
\begin{align*}
    \epsilon_i^\text{idle} &= \epsilon_i^\text{relaxation} + \epsilon_i^\text{dephasing}\\
    &= 1 - \exp\left(\frac{-t}{T^{(i)}_1}\right) + \frac{1}{2}\exp\left(\frac{-t}{T^{(i)}_1}\right)\left(1 - \exp\left(-t \left(\frac{1}{\min(T^{(i)}_1,T^{(i)}_2)} - \frac{1}{T^{(i)}_1}\right)\right)\right).
\end{align*}
Each time step of SWAP gates contributes 3 CNOT idle errors to each idle abstract qubit. Additionally, for this extended error model, we adjust the CNOT errors to split evenly between abstract qubits. This is because there is a movement variable for both qubits involved in the swap operation. To keep idle errors and gate errors proportionate to the physical device, we avoid double counting the CNOT gate errors $\epsilon_\alpha^\text{CNOT}$ by splitting them between the two corresponding movement variables to get an adjusted error $\epsilon_\alpha^v$ which can be used to calculate the adjusted weight for each of the variables. This is done by equating the success probabilities~$(1 - \epsilon_\alpha^\text{CNOT}) = (1 - \epsilon_\alpha^v) * (1 - \epsilon_\alpha^v)$ and solving to get
\begin{equation}
    \epsilon_\alpha^v = 1 - \sqrt{1 - \epsilon_\alpha^\text{CNOT}}.
\end{equation}
The corresponding final objective function for the extended noise model is
\begin{equation}
    C = -\sum_{\alpha\in \{\text{SWAPs}\}} 3\log(1-\epsilon_\alpha^v) - \sum_{i\in \{\text{idle}\}} 3\log(1-\epsilon_i^\text{idle}),\label{eq:full-cost}
\end{equation}
where the two sums are over the sets of SWAP gates and idle abstract qubits for all time steps of the solution.

The MQPF algorithm can be adjusted to prioritise optimisation of the accumulated error by considering solutions with higher SWAP depths, which may involve a trade-off between the optimisation quality of the SWAP depth and error objectives.

We will describe how the BILP is constructed in the following sections.

\subsubsection{Multi-Commodity Flow Problem}

Our approach to finding valid paths on a time-expanded graph draws inspiration from the multi-commodity flow problem. We will begin by constructing the linear program for multi-commodity flow on the time-expanded graph as a base, then modify it to obtain the BILP for the MQPF problem. The multi-commodity flow problem typically represents scenarios in network flow where multiple flows, each corresponding to a distinct commodity, need routing through a network. A classical example is the telecommunication routing problem, where messages are routed through a network to minimise costs associated with transmission line variables such as packet delay and bandwidth utilisation~\cite{mahey1998new}. In such models, network nodes represent traffic origin and destination stations and edges represent transmission lines. The objective is to route a specified amount of messages from the origins to the destinations without exceeding the capacity of any transmission line, all while minimising the total cost.

For the quantum hardware layout, we adopt a similar approach but apply it to a time-expanded graph. This adaptation can be modelled through the following linear program.
\newline{}
\newline{}
\noindent Objective Function:
\begin{equation}
    \min {\sum_{k\in \{1,\ldots K\}}\sum_{t\in \{1,\ldots T\}}\sum_{(i,j)\in E} c_{ij}x_{ij,t}^k},
\end{equation}
where $\{1,\ldots K\}$ are the commodities (teams of qubits), $\{1,\ldots T\}$ are the time steps, $(i,j)\in E$ are the edges in the hardware graph $G$ connecting nodes $i$ and $j$ from the set of nodes $V$ of $G$, $c_{ij}$ is the cost weight of edge $(i,j)$, and $x_{ij,t}^k$ is the flow of commodity $k$ through edge $(i,j)$ from $i$ at time~$t-1$ to $j$ at time $t$.
\newline{}
\newline{}
\noindent Constraints:
\begin{itemize}
    \item[(1)] \textbf{Conservation of flow:} The flow of a commodity entering a time-expanded graph node must be equal to the flow exiting it in the next time step.\\
    $\forall k \in \{1,\ldots K\}, \;\; \forall t \in \{1,\ldots T-1\},\;\; \forall i\in V, \;\; \sum_{j\in V} x_{ji,t}^k - \sum_{j\in V}  x_{ij,t+1}^k = 0$,\\
    $\forall k \in \{1,\ldots K\}, \;\;\; \forall i\in \boldsymbol{S}_k,\;\;\; x_{I_k i} - \sum_{j\in V} x_{ij,1}^k = 0$,\\
    $\forall k \in \{1,\ldots K\}, \;\;\;\forall i\in \boldsymbol{D}_k,\;\;\;\sum_{j\in V} x_{ji,t_{T-1}}^k - x_{i F_k} = 0,$
    \item[(2)] \textbf{Edge flow capacity:} The total amount of flow on each edge can never be more than its assigned maximum flow capacity.\\
    $\forall t \in \{1,\ldots T\}, \;\;\; \forall (i,j) \in E, \;\;\; \sum_{k\in K} x_{ij,t}^k \leq u_{ij}$,
    \item[(3)] \textbf{Maximum flow for source and destination edges:} Ensures that all of each commodity is routed in any valid solution. \\
    $\forall k\in \{1,\ldots K\}, \;\;\;\forall i\in \boldsymbol{S_k},\;\;\; x_{I_k i} = u_{I_k i},$\\
    $\forall k\in \{1,\ldots K\}, \;\;\;\forall i\in \boldsymbol{D_k},\;\;\; x_{i F_k} = u_{iF_k},$
    \item[(4)] \textbf{Real variables:} The amount of commodity flow through edges is continuous, not discrete.\\
    $\forall k \in \{1,\ldots K\},\;\;\; \forall t\in \{1,\ldots T\},\;\;\; \forall (i,j)\in E,\;\;\;  x_{ij,t}^k\in \mathbb{R}$,
\end{itemize}
where the variables $x_{I_k i}$ and capacities $u_{I_k i}$ correspond to flows and max flows from the attached special source node $I_k$ to source nodes $i\in \boldsymbol{S_k}$ and the variables $x_{iF_k}$ and capacities $u_{i F_k}$ correspond to flows and max flows from destination nodes $i\in\boldsymbol{D}_k$ to the attached special destination node $F_k$. The capacities $u_{ij}$ represent the maximum allowable flow through edge $(i,j)\in E$ of $G$ at any time step.

\subsubsection{Adapting Multi-Commodity Flow to MQPF}\label{sec:adapting-multi-commodity-flow}

\begin{figure}
     \centering
     \includegraphics[scale=1.2]{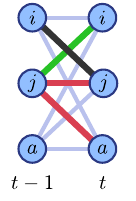}
    \caption{A graphic representing Constraint (6) introduced in Adjustment 4 of Section~\ref{sec:adapting-multi-commodity-flow} which enforces swap-based movement when adapting the linear program from multi-commodity flow to multi-qubit pathfinding. If the abstract qubit at location $i$ time step $t-1$ moves to location $j$ time step $t$ (shown in black), then the abstract qubit at location $j$ time step $t-1$, in this example, cannot move to locations $j$ nor $a$ at time step $t$ (shown in red). Therefore, the only movement possible for abstract qubit at location $j$ time step $t-1$ is to location $i$ time step $t$ (shown in green). \label{fig:time-expand-constraint}}
\end{figure}
The multi-commodity flow linear program can be modified to mimic the behaviour of teams of abstract qubits traversing a hardware graph via SWAP gates. The following four adjustments can be made to achieve this: 
\begin{itemize}
    \item[1.] \textbf{Modification of Variables to Binary:} The variables are changed to binary, modifying Constraint (4) as:
    \begin{itemize}
        \item[(4)] Binary variables: \\
        $\forall k \in \{1,\ldots K\},\;\;\; \forall t\in \{1,\ldots T\},\;\;\; \forall (i,j)\in E,\;\;\;  x_{ij,t}^k\in \{0, 1\}$.
    \end{itemize}
    With this adjustment, each variable $x_{ij,t}^k$ now decides whether an abstract qubit from team $k$ is traversing the edge (corresponding to a two-qubit swap or idle at location) in the time-expanded graph or not.
    \item[2.] \textbf{Setting Flow Capacities to One:} To enforce that no more than one abstract qubit traverses any edge in the time-expanded graph at a time, all flow capacities are set to one:
    \begin{itemize}
        \item $\forall (i,j)\in E, \;\;\;u_{ij} = 1$,
        \item $\forall k \in \{1, \ldots K\}, \;\;\; \forall j \in \boldsymbol{S}_k, \;\;\; u_{I_k j} = 1$,
        \item $\forall k \in \{1, \ldots K\}, \;\;\; \forall i \in \boldsymbol{D}_k,\;\;\; u_{i F_k}=1$.
    \end{itemize}
    \item[3.] \textbf{Enforcing Exclusivity of Location:} A fifth constraint is introduced to limit each location in the time-expanded graph to a maximum of one abstract qubit per time step:
    \begin{itemize}
        \item[(5)] Exclusivity of location:\\
        $\forall t\in\{1,\ldots T\}, \;\;\; \forall i\in V,\;\;\;\sum_{k=1}^K\sum_{j\in V} x_{ji,t}^{k} \leq 1$.
    \end{itemize}
    \item[4.] \textbf{Enforcing Swap-Based Movement:} A sixth constraint is introduced to enforce the swapping behaviour of abstract qubits:
    \begin{itemize}
        \item[(6)] Swap-based movement:\\
        $\forall t\in\{1,\ldots T\}, \; \forall (i,j)\in E,\;\sum_{k=1}^K x_{ij,t}^{k} + \sum_{k=1}^K\sum_{l\in V, \text{s.t. }l\neq i } x_{jl,t}^{k} \leq 1$.
    \end{itemize}
\end{itemize}
To help better understand Adjustment 4, a graphical representation of the newly added Constraint (6) is shown in Figure~\ref{fig:time-expand-constraint}. The constraint ensures that if an abstract qubit traverses from location $i$ to $j$ at time step $t$, then any qubit at $j$ in the previous time step can only move to location $i$. To see how this works in more detail, we can look at each term separately. The first summation term in the inequality is over the flows along edge $(i,j)$ which maximises at~1 due to the total number of abstract qubits that can travel through an edge being limited by a capacity of~1. The second summation term is over all flows from location $j$ that do not go to location $i$ in time step $t$. The sum is always a maximum of~1 since all variables~$x_{jl,t}^{k}$ in the sum correspond to possible movement options for an abstract qubit at location $j$ and there can only be one abstract qubit at any location at any time (as per Constraint (5)). Adding the two summation terms together and requiring them to be no more than 1 forces a maximum of only one term to be equal to 1. Thus, both of the following cannot be simultaneously true in the same time step: that abstract qubit at $i$ moves to $j$ and abstract qubit at $j$ moves to somewhere that is not $i$, hence enforcing the swap behaviour.

\begin{figure}
     \centering
     \subfloat[][Example hardware graph]{
         \centering
         \includegraphics[scale=2.4]{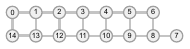}
     }
     \qquad
     \subfloat[][Example optimal solution paths]{
     \centering
     \includegraphics[width=0.60\textwidth]{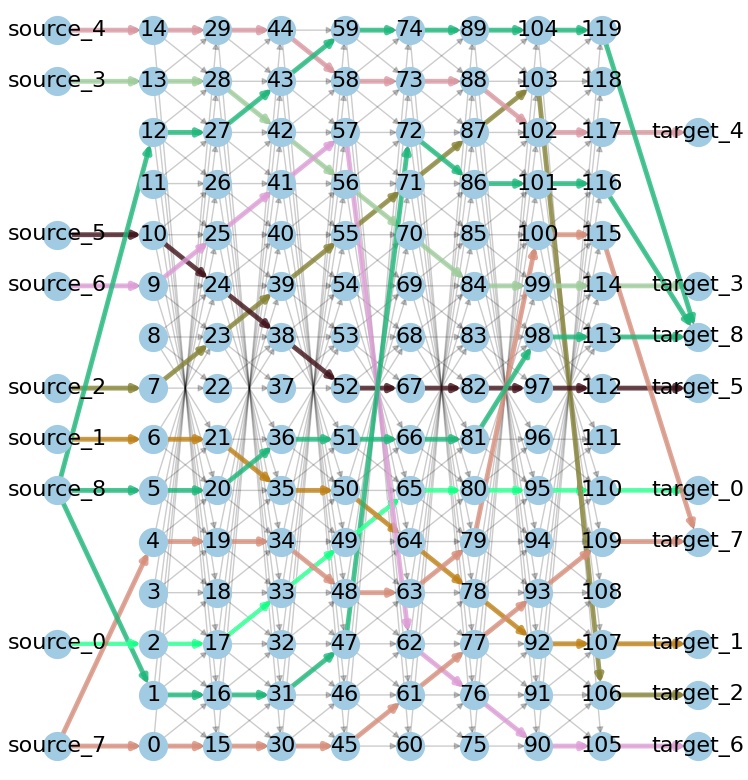}
     }
        \caption[An example multi-qubit pathfinding solution for a random problem instance on the 15-qubit \textit{ibmq\_16\_melbourne} device with a time-depth of 7.]{An example multi-qubit pathfinding solution for a random problem instance on the 15-qubit \textit{ibmq\_16\_melbourne} device. \textbf{(a)} The \textit{ibmq\_16\_melbourne} device qubit layout. \textbf{(b)} Each attached special source node, labelled `source\_$i$', is assigned a colour corresponding to a team $i$ of abstract qubits and directed edges connect to corresponding source nodes. The attached special destination nodes for team $i$ are labelled `target\_$i$' and have edges connecting to the corresponding destination nodes. The coloured paths of edges correspond to paths through time for each abstract qubit in the optimal SWAP-depth solution to the example multi-qubit pathfinding problem instance. \label{fig:example-pathfinding-solution}}
\end{figure}

\subsubsection{The final BILP for the MQPF problem}
The BILP used to search for a solution at each time step during the iterative time expansion of our MQPF algorithm can be summarised as follows.
\newline{}
\newline{}
\noindent Objective Function:
\begin{equation}
        \min {\sum_{k\in \{1,\ldots K\}}\sum_{t\in \{1,\ldots T\}}\sum_{(i,j)\in E} c_{ij}x_{ij,t}^k},
\end{equation}
where $\{1,\ldots K\}$ are the teams of abstract qubits, $\{1,\ldots T\}$ are the time steps, $(i,j)\in E$ are the qubit-pair edges in the hardware graph $G$, $c_{ij} := -3\log(1-\epsilon_{ij})$ is the cost for a SWAP gate to be applied to qubit pair $(i,j)$ (specified in Equation~\ref{eq:objective_value}), and $x_{ij,t}^k$ specifies whether a team $k$ abstract qubit traverses qubit-pair $(i,j)$ from qubit $i$ at time~$t-1$ to qubit~$j$ at time $t$.
\newline{}
\newline{}
\noindent Constraints:
\begin{itemize}
    \item[(1)] \textbf{Conservation of flow:} The number of abstract qubits of a team swapping to or idling at a location, must be equal to the number swapping away from or idling at that location in the next time step.\\
    $\forall k \in \{1,\ldots K\}, \;\; \forall t \in \{1,\ldots T-1\},\;\; \forall i\in V, \;\; \sum_{j\in V} x_{ji,t}^k - \sum_{j\in V}  x_{ij,t+1}^k = 0$,\\
    $\forall k \in \{1,\ldots K\}, \;\;\; \forall i\in \boldsymbol{S}_k,\;\;\; x_{I_k i} - \sum_{j\in V} x_{ij,1}^k = 0$,\\
    $\forall k \in \{1,\ldots K\}, \;\;\;\forall i\in \boldsymbol{D}_k,\;\;\;\sum_{j\in V} x_{ji,t_{T-1}}^k - x_{i F_k} = 0,$
    \item[(2)] \textbf{Edge flow capacity:} There can only be up to one abstract qubit swapping between a qubit pair or idling at a location at any one time.\\
    $\forall t \in \{1,\ldots T\}, \;\;\; \forall (i,j) \in E, \;\;\; \sum_{k\in K} x_{ij,t}^k \leq 1$,
    \item[(3)] \textbf{Maximum flow for source and destination edges:} Ensures that any valid solution has all source locations start with an abstract qubit of the corresponding team and similarly all destination locations end with an abstract qubit of the corresponding team.\\
    $\forall k\in \{1,\ldots K\}, \;\;\;\forall i\in \boldsymbol{S_k},\;\;\; x_{I_k i} = 1,$\\
    $\forall k\in \{1,\ldots K\}, \;\;\;\forall i\in \boldsymbol{D_k},\;\;\; x_{i F_k} = 1,$
    \item[(4)] \textbf{Binary variables:} An abstract qubit is either traversing an edge (corresponding to a two-qubit swap or idle at location) or it is not.\\
    $\forall k \in \{1,\ldots K\},\;\;\; \forall t\in \{1,\ldots T\},\;\;\; \forall (i,j)\in E,\;\;\;  x_{ij,t}^k\in \{0, 1\}$,
    \item[(5)] \textbf{Exclusivity of location:} Each location can only have up to one abstract qubit at a time.\\
    $\forall t\in\{1,\ldots T\}, \;\;\; \forall i\in V,\;\;\;\sum_{k=1}^K\sum_{j\in V} x_{ji,t}^{k} \leq 1$,
    \item[(6)] \textbf{Swap-based movement:} Enforces a swap behaviour between qubit pairs by disallowing both having an abstract qubit move from $i$ to $j$ and an abstract qubit move from $j$ to somewhere other than $i$ in the same time step.  \\
    $\forall t\in\{1,\ldots T\}, \; \forall (i,j)\in E,\;\sum_{k=1}^K x_{ij,t}^{k} + \sum_{k=1}^K\sum_{l\in V, \text{s.t. }l\neq i } x_{jl,t}^{k} \leq 1$.
\end{itemize}

The program can be further tailored to accommodate scenarios where destination nodes outnumber source nodes, and where multiple teams may share some destination nodes. Such flexibility is beneficial when the precise assignment of destination nodes is part of the optimisation problem, allowing for some leeway in these assignments. This can be implemented by removing the requirement for maximum flow at all destination edges (previously specified in Constraint (3)). By maintaining flow conservation, along with maximum flow constraints at all source edges, the program ensures that each abstract qubit navigates through the time-expanded graph to the correct special destination node $F_k$ associated with its team. In addition to enhancing the program's flexibility, it can also reduce the optimal SWAP depth and hence computation time by broadening the range of destination options available for abstract qubits.

\section{Results}\label{sec:results}

\subsection{MQPF on Physical Qubit Layouts}\label{sec:mqp-results}

The multi-qubit pathfinding (MQPF) algorithm described above was programmed in Python. The script inputs a hardware graph, abstract qubit teams, and their corresponding source and destination locations. It then automatically time expands the hardware graph incrementally for increasing depth and generates the corresponding binary integer linear programs (BILPs). Each program is solved using Gurobi commercial optimiser which specialises in solving mathematical programs~\cite{gurobi}. Computations were performed on an Intel Core i5-6500 CPU 3.2 GHz running Windows 10 with 16GB RAM. Instances exceeding 1000 seconds of computation time were recorded and abandoned. When a solution is found, it can be used to construct the paths through time for each of the abstract qubits. An example of a solution for a random problem instance on the \textit{ibmq\_16\_melbourne} device is shown in Figure~\ref{fig:example-pathfinding-solution}.

In our implementation, we improve the algorithm's computational efficiency using the following two optimisations:
\begin{itemize}
    \item[1.] Trim impossible variables.
    \item[2.] Presolve using faster algorithms to obtain SWAP-depth lower bounds.
\end{itemize}
In Optimisation~1, all variables corresponding to edges that are impossible to reach for the associated variable's team are removed. This is done with a sweep forwards in time to remove variables for edges that abstract qubits cannot reach and then a sweep backwards in time to remove variables for edges that abstract qubits cannot use, because if they did, they would not be able to reach any of their destination nodes by the final time step. In Optimisation~2, each iteration of the time expansion is first solved using a fast algorithm with a solution that is a lower-bound of the SWAP-depth solution. The same MQPF algorithm can be used to achieve this when all abstract qubits are assigned to the same team, since the computation speed can become orders of magnitude faster. Another fast lower-bound algorithm approach is to use Dijkstra's pathfinding algorithm to obtain the maximum length over teams, where each team length is calculated as the minimum path distance over individual paths from source nodes to target nodes in the same team. Then, once a solution is found, the time expansion procedure continues but with the qubits assigned to their usual teams and the depth starting at the depth of the found solution. Optimality is not lost since the optimal SWAP-depth for multiple teams is always greater than or equal to that of a single team for the same source and destination locations on the hardware graph. 

\begin{figure}
     \centering
     \includegraphics[width=0.46\textwidth]{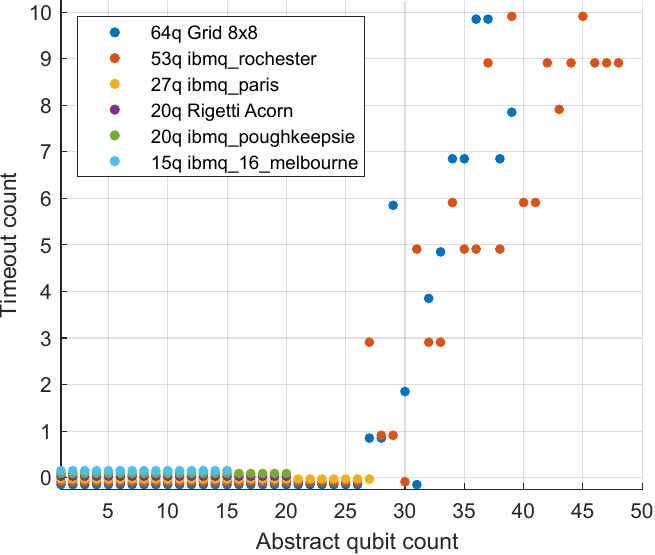}
        \caption{The number of timeouts that occurred when multi-qubit pathfinding numbers of independent abstract qubits (in their own teams) on the physical device layouts shown in Figure~\ref{fig:hardware-graphs}. Each data point is the number of random instances out of 10 that exceed 1000 seconds of computation time. Note that all instances that were above 39 independent abstract qubits on the 64-qubit 8x8 grid and above 48 on the \textit{ibmq\_rochester} device timed out. All instances that time out are abandoned and ignored in the results. There is a small offset along the vertical axis of each device for the purpose of displaying overlapping data points. }\label{fig:timeouts-vs-agent_counts}
\end{figure}

\begin{figure}
     \centering
     \subfloat[\label{fig:average_makespan-vs-agent_count-with_timeout}]{
         \centering
         \includegraphics[width=0.46\textwidth]{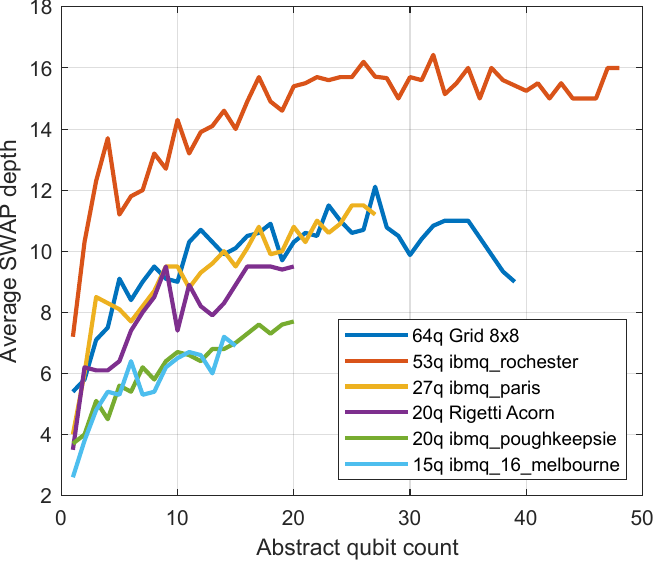}
     }
     \qquad
     \subfloat[\label{fig:instance_depth-vs-agent_count-with_timeout}]{
         \centering
         \includegraphics[width=0.46\textwidth]{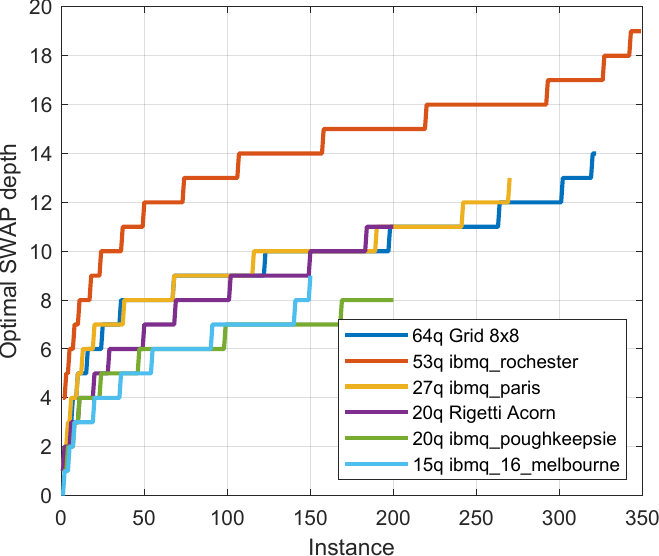}
     }
        \caption[Optimal SWAP depth solutions for multi-qubit pathfinding (MQPF) abstract qubits on the 6 hardware graphs.]{Optimal SWAP depth solutions for multi-qubit pathfinding (MQPF) abstract qubits on the 6 hardware graphs displayed in Figure~\ref{fig:hardware-graphs}, found using the MQPF algorithm presented in Section~\ref{sec:mqpf_algorithm}. There are 10 random instances for each number of abstract qubits, starting at one and ending at the number of physical qubits in the corresponding hardware graph. Each abstract qubit is in its own team. Problem instances that exceed 1000 seconds are considered timed out and are ignored, with the number of instances timed out for each number of abstract qubits shown in Figure~\ref{fig:timeouts-vs-agent_counts}. \textbf{(a)}~The average optimal SWAP depths for each number of abstract qubits. The SWAP depths appear to plateau as the number of abstract qubits increases, which could be due to the instances with higher SWAP depths being more likely to timeout. \textbf{(b)}~Individual optimal SWAP depths for all instances. The instance numbers shown for each line of data are labelled based on sorting the data in ascending order with respect to the y-axis values, they are not unique to particular instances.}\label{fig:makespan-vs-abstract-qubit-count}
\end{figure}

We benchmark the overall performance of the pathfinding algorithm on a variety of hardware graphs shown in Figure~\ref{fig:hardware-graphs} by solving random problem instances. In all of the following computations, instances that exceeded 1000 seconds with respect to the single CPU machine were ignored in the results, the total number of timeouts that occurred for each problem size is shown in Figure~\ref{fig:timeouts-vs-agent_counts}. First we focus on the cases where each abstract qubit is in its own team and measure the optimal SWAP depths as shown in Figure~\ref{fig:makespan-vs-abstract-qubit-count}. The first plot shows the average solution SWAP depth over 10 random instances for increasing numbers of abstract qubits (starting at one and ending at the number of qubits in the corresponding hardware graph), while the second plot shows solution SWAP depths for all of the random instances sorted in order of increasing optimal SWAP depth. Instances are randomised by sampling a random set of locations equal to the number of abstract qubits for the starting nodes and again for the ending nodes, where the index of elements in the sets correspond to the index of the abstract qubits. When comparing the 53-qubit Rochester device and 64-qubit 8x8 grid, there is a considerable reduction in solution SWAP depth for the grid, even though it consists of more qubits. This is likely due to its connectivity, the average number of qubits each qubit is connected to is higher for the grid than for the Rochester device. This allows the abstract qubits to find more direct paths to their destinations. At about 27 abstract qubits, the SWAP depth appears to flatten out and then decrease for increasing problem size. This is likely due to the timed out instances not being included in the calculation of the mean, causing a bias towards shorter depth solutions, since the SWAP depth is a significant factor in the overall runtime of the algorithm. Considering the additional variables and constraints included from each added abstract qubit, the algorithm is going to reach the timeout at shorter depths for larger problem instances.

\begin{figure}
     \centering
     \subfloat[\label{fig:average_total_costs-vs-layouts-agent_count}]{
         \centering
          \includegraphics[width=0.46\textwidth]{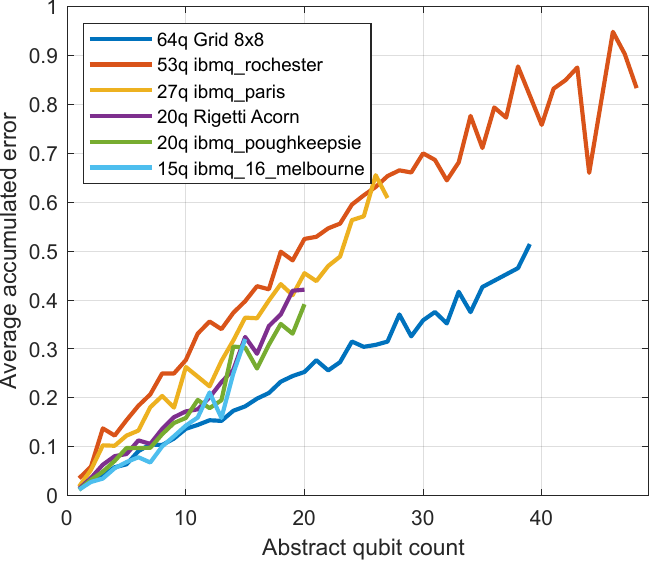}
     }
     \qquad
     \subfloat[\label{fig:instance_total_costs-vs-layouts-agent_count}]{
         \centering
         \includegraphics[width=0.46\textwidth]{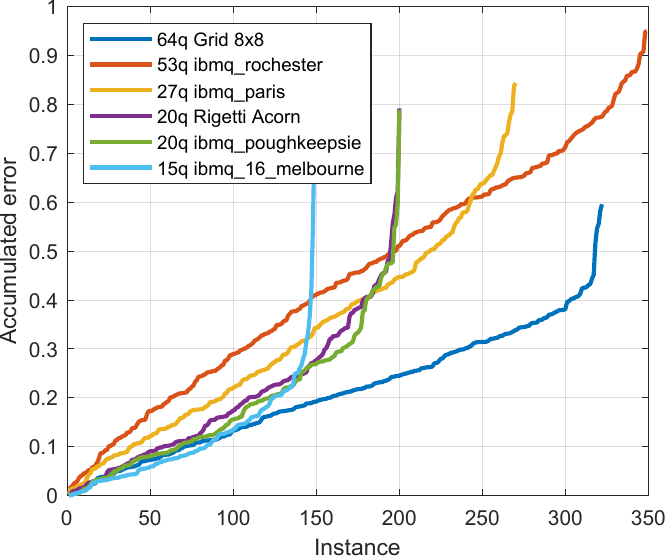}
     }
        \caption[Optimised accumulated error rates in optimal SWAP-depth solutions for multi-qubit pathfinding (MQPF) abstract qubits on the 6 hardware graphs.]{Optimised accumulated error rates in optimal SWAP-depth solutions for multi-qubit pathfinding (MQPF) abstract qubits on the 6 hardware graphs displayed in Figure~\ref{fig:hardware-graphs}, found using the MQPF algorithm presented in Section~\ref{sec:mqpf_algorithm}. There are 10 random instances for each number of abstract qubits. The CNOT errors are randomised for each instance from a truncated normal distribution on a log scale with a mean of 0.001 and a log$_{10}$ error variance of 0.5. Each abstract qubit is in its own team. Problem instances that exceed 1000 seconds are considered timed out and are ignored, with the number of instances timed out for each number of abstract qubits shown in Figure~\ref{fig:timeouts-vs-agent_counts}. \textbf{(a)} The average accumulated error rates for each number of abstract qubits. The variance of the accumulated errors increases due to there being more timeouts for higher numbers of abstract qubits. \textbf{(b)} Individual accumulated error rates for all instances. The instance numbers shown for each line of data are labelled based on sorting the data in ascending order with respect to the y-axis values, they are not unique to particular instances.}\label{fig:error-vs-qubit-count-comparisons}
\end{figure}

Next, we measure the optimised accumulated error rates for the time-expanded graphs at the optimal SWAP depths, shown in Figure~\ref{fig:error-vs-qubit-count-comparisons}. The average accumulated error over 10 random instances for each number of abstract qubits is shown in Figure~\ref{fig:average_total_costs-vs-layouts-agent_count}. The accumulated error for all of the random instances sorted in order of increasing error rate is shown in Figure~\ref{fig:instance_total_costs-vs-layouts-agent_count}. The CNOT errors (where 3 CNOT gates are used to implement a SWAP gate) assigned to physical qubit connections are randomly chosen using a log scale truncated normal distribution with a CNOT error mean of~0.001 and a log$_{10}$ variance of~0.5. A log scale normal distribution is used because it roughly matches the distributions of CNOT error rates found in the \textit{ibmq\_melbourne} and \textit{ibmq\_rochester} devices, although both with a~$\log_{10}$ standard deviation of approximately~0.165. The larger variance was chosen in the experiments to more clearly compare the differences between the devices. The~64-qubit 8x8 grid showed a considerable improvement over the other devices even though it has more physical qubits. This is likely due to the following two reasons. The first is that the 8x8 grid results in shallower optimal SWAP depths than some smaller devices such as the \textit{ibmq\_rochester} device, shown in Figs.~\ref{fig:average_makespan-vs-agent_count-with_timeout} and~\ref{fig:instance_depth-vs-agent_count-with_timeout}. Shallower SWAP depths typically use fewer SWAP gates, hence fewer sources of error. The second reason for the 8x8 grid's performance is that with the higher connectivity, there are more qubit neighbour options for each abstract qubit to swap with, thus providing more opportunities to swap along connections with lower physical error rates. 

\begin{figure}
     \centering
     \includegraphics[width=0.46\textwidth]{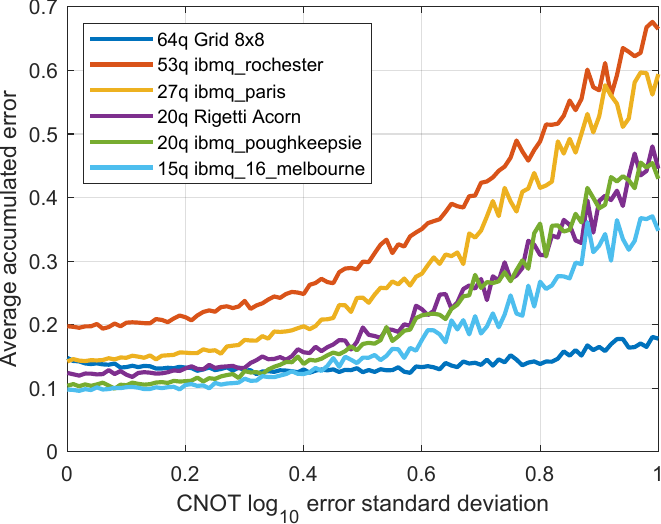}
        \caption[Average accumulated error for pathfinding random instances of 10 abstract qubits with random CNOT error rates for increasing variance on the 6 hardware graphs.]{Average accumulated error for pathfinding random instances of 10 abstract qubits with random CNOT error rates for increasing variance on the 6 hardware graphs displayed in Figure~\ref{fig:hardware-graphs}. Each data point is averaged over 100 random problem instances with CNOT errors sampled from truncated normal distributions on a log scale with a mean of 0.001 and specified log$_{10}$ error standard deviations. The log$_{10}$ standard deviations range from 0 to 1 increasing in increments of 0.01. Problem instances that exceed 1000 seconds are considered timed out and are ignored in the results. A log$_{10}$ standard deviation of 1 corresponds to 1 order of magnitude. }\label{fig:average_total_costs-vs-error_stds-comparison}
\end{figure}

The variance used for the distribution of CNOT-gate errors on the hardware graph can influence the accumulated error rates of optimised qubit paths. To investigate its impact, we measured the average optimal accumulated error for pathfinding random instances of~10 independent abstract qubits on hardware graphs that are assigned random CNOT error rates sampled from distributions of increasing log variance. The results are shown in Figure~\ref{fig:average_total_costs-vs-error_stds-comparison}. Each data point is the average of 100 random problem instances where each instance has newly randomised CNOT-errors on the hardware graph. The results show that accumulated error rates tend to increase for increasing CNOT-error variance. This is likely due to the multiplicative nature of gate errors making high error gates far more impactful than low error gates. However, on the 64-qubit 8x8 grid, the accumulated error rates initially decrease until a CNOT $\log_{10}$ error standard deviation of approximately 0.45, where it begins to increase. This initial decrease is likely due to the higher connectivity providing more opportunities to avoid higher error-rate CNOT gates, and a lower variance means that whatever higher error-rate gates that are included tend to be less extreme.

\begin{figure}
     \centering
     \subfloat[\label{fig:average_runtimes-vs-agent_count-comparison}]{
         \centering
          \includegraphics[width=0.45\textwidth]{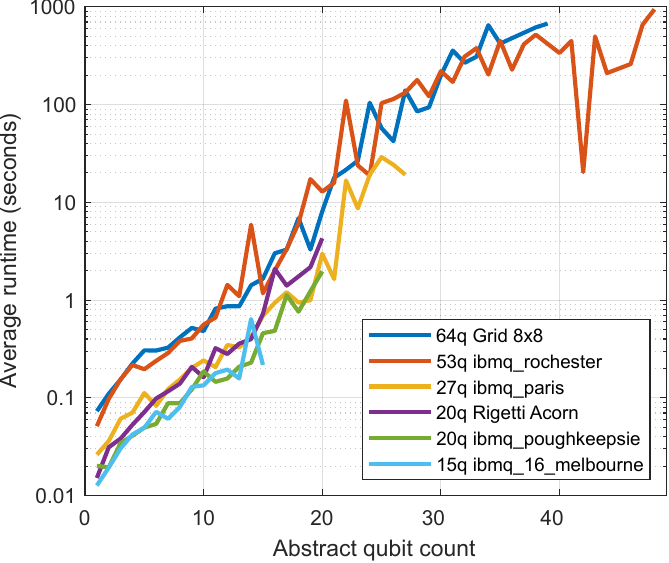}
     }
     \qquad
     \subfloat[\label{fig:instance_runtimes-vs-agent_count-comparison}]{
         \centering
         \includegraphics[width=0.47\textwidth]{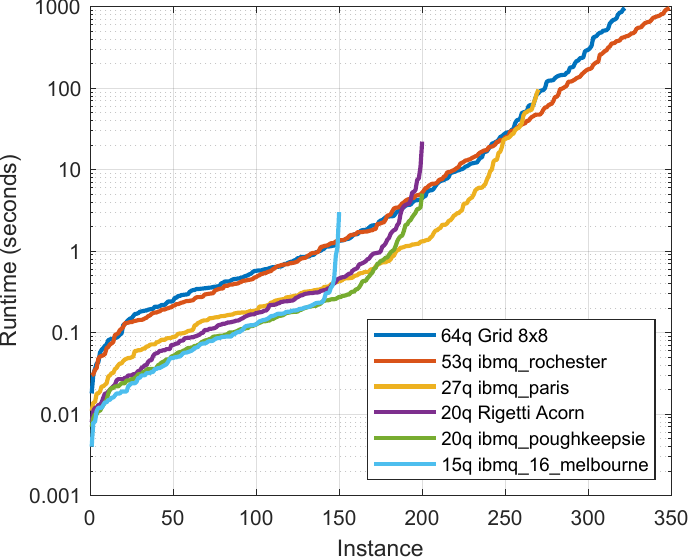}
     }
        \caption[Algorithm computational runtimes for multi-qubit pathfinding abstract qubits on the 6 hardware graphs.]{Algorithm computational runtimes for multi-qubit pathfinding abstract qubits on the 6 hardware graphs displayed in Figure~\ref{fig:hardware-graphs}. There are 10 random instances for each number of abstract qubits. Each abstract qubit is in its own team. Problem instances that exceed 1000 seconds are considered timed out and are ignored, with the number of instances timed out for each number of abstract qubits shown in Figure~\ref{fig:timeouts-vs-agent_counts}. The runtimes scale with the number of variables and constraints in the binary integer linear program generated for the solutions. \textbf{(a)} The average runtime for each number of abstract qubits. \textbf{(b)} Individual runtimes for all instances. The instance numbers shown for each line of data are labelled based on sorting the data in ascending order with respect to the y-axis values, they are not unique to particular instances.}
\end{figure}

The average runtime over 10 random instances was measured for increasing numbers of abstract qubits and the results are plotted in Figure~\ref{fig:average_runtimes-vs-agent_count-comparison}. The runtimes for all of the instances are plotted in Figure~\ref{fig:instance_runtimes-vs-agent_count-comparison}, sorted in ascending order with respect to runtime. The runtime generally increases for increasing numbers of variables and constraints in the BILP corresponding to the problem instance. Thus, solutions for larger hardware graphs typically take longer to compute. Although, hardware graphs with higher connectivity typically result in lower SWAP depth solutions, leading to shorter computation times. This can be seen, for example, when comparing results for the 64-qubit 8x8 grid against the 53-qubit \textit{ibmq\_rochester} device. The grid has comparable runtimes to the \textit{ibmq\_rochester} device even though it contains $20\%$ more physical qubits and $93\%$ more edges, adding variables and constraints to the BILP for the same time-expanded depth.

\begin{figure}
     \centering
     \includegraphics[width=0.5\textwidth]{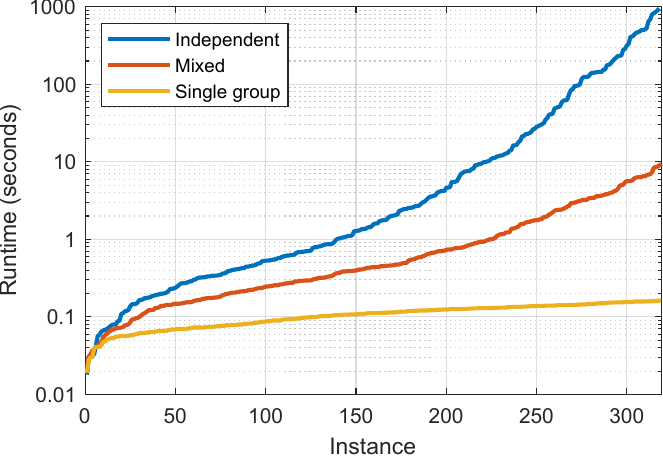}
        \caption{Computational runtimes for three methods of assigning abstract qubits to teams on the 64-qubit 8x8 grid hardware graph. The three methods used were (where $N$ is the number of abstract qubits): independent, where each abstract qubit is in its own team ($N$ teams); mixed, where the set of abstract qubits are randomly assigned to a random number of teams (an average of $N/2$ teams); and single group, where all abstract qubits are in a single team ($1$ team). There are~10 random instances for each number of abstract qubits from~1 to~64. The instance numbers shown for each line of data are labelled based on sorting the data in ascending order with respect to the y-axis values, they are not unique to particular instances.
        }\label{fig:grid8x8-benchmarking}
\end{figure}

Finally, since the number of variables and constraints scale with the number of teams, we investigate the effect that the number of teams has on computational runtimes. The runtimes for 10 random instances on the 64-qubit 8x8 grid hardware graph for increasing numbers of abstract qubits, $N$, are plotted in Figure~\ref{fig:grid8x8-benchmarking}. Three cases were compared: independent abstract qubits that are each in their own team ($N$ teams); abstract qubits that are randomly assigned to a random number of teams (an average of $N/2$ teams); and a single team consisting of all abstract qubits (1 team). The algorithm appears to scale significantly better with fewer numbers of teams. The data for mixed teams is roughly centred between independent and single group teams on a log scale, this is an indication that the runtime is increasing roughly exponentially with the number of teams, as we would expect. The SWAP depth is typically larger in solutions with more teams since conflicts between teams are more frequent and there is less sharing of target destinations among qubits, that is, tighter restrictions on which qubits need to go where.

\subsection{Comparison with the SMT-CBS Algorithm} \label{sec:comparison_with_smtcbs}
We extend the previous experiments by comparing the MQPF algorithm against the Satisfiability Modulo Theories and Conflict-Based Search (SMT-CBS) algorithm~\cite{surynek2019multi, surynek2019conflict} for multi-qubit pathfinding. The SMT-CBS algorithm uses a high level CBS layer to search a constraint tree in a breadth first search manner, where each node is a set of collision avoidance constraints, while a low level SMT-based process is used on each node of the constraint tree to determine if a conflict free solution exists. The code for the SMT-CBS algorithm was obtained from the boOX github repository~\cite{boox}. We used the TSWAP solver and an algorithm called non-refined SAT (NRF-SAT) which includes a compilation scheme that replaces path-consistency constraints in SMT-CBS with a post-processing step to extract paths. This implementation of the SMT-CBS algorithm exactly optimises the sum of time steps for each agent to reach their destination, rather than the makespan. This approach is effective at reducing the total SWAP count, however is not always optimal. The SMT-CBS algorithm is compared with 3 variations of the MQPF algorithm introduced in this work: optimal MQPF, near-optimal MQPF, and feasible MQPF. Optimal MQPF uses the Gurobi optimiser with default settings which sets MIPFocus to 0 (a balanced approach to find optimal solutions), MIPGap to 0.0001, and MIPGapAbs to $10^{-10}$. The near-optimal MQPF sets MIPFocus to 1 (to find feasible solutions quickly), MIPGap to~0.08, and MIPGapAbs to~0.08. The feasible MQPF sets MIPFocus to 1 and SolutionLimit to 1 so that the algorithm stops at the first feasible solution it finds.

For these experiments, we use the objective function for the extended noise model that includes idle errors described in Equation~\ref{eq:full-cost}. The CNOT idle time for the idle error calculations was chosen based on the average CNOT gate time on an IBM Quantum Heron device, which was~79~ns. We generated 10 random problem instances for each number of abstract qubits assessed. Each one has an error map, which includes CNOT errors as well as qubit $T_1$ and $T_2$ times that were randomly generated using parameters chosen to match the current settings of IBM Quantum Heron devices. For two-qubit gates, errors were sampled from a truncated normal distribution with 0.007 mean and 0.3115 log10 standard deviation bounded by $\epsilon^\text{CNOT}_\alpha \in [0.0018, 0.0876]$. The $T_1$ and $T_2$ times are sampled from truncated normal distributions, such that, the $T_1$ distribution mean is 176 ms with $\sigma=69$ bounded by $T_1 \in [3, 310]$, and the $T_2$ distribution has mean 140 ms with $\sigma=71$ bounded by $T_2 \in [6, 321]$. 

\begin{figure}
     \centering
     \includegraphics[width=1\textwidth]{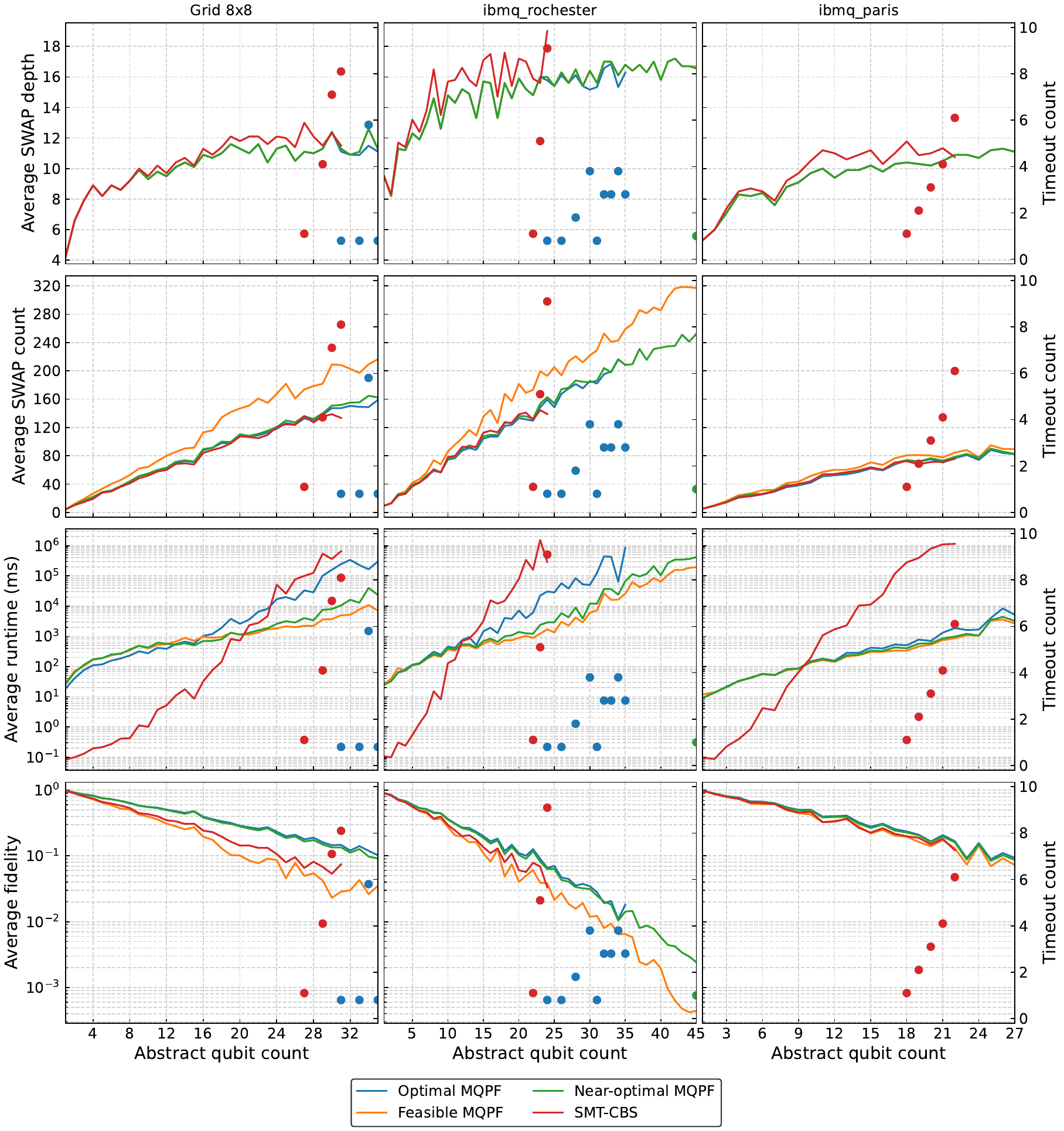}
        \caption{Comparison of algorithms on the 64q 8x8 Grid, 53q \textit{ibmq\_rochester} and 27q \textit{ibmq\_paris} layouts. There are~10 random instances for each number of abstract qubits. The fidelity is calculated from the accumulated errors of both SWAP gates and idling qubits. The y-axis values are the mean over completed instances for each abstract qubit count. The displayed individual data points are the number of instances out of 10 that timed out, exceeding 3600 seconds of computation time. These points have a small offset along the vertical axis for each algorithm to help distinguish any overlapping data points. 
        }\label{fig:comparison-summary}
\end{figure}

The hardware configuration used for experiments in the previous section was no longer available to gather the data for this section. Instead we used an AMD Ryzen 7 5700X 8-Core CPU 3.4 GHz running Windows Subsystem for Linux (WSL) in Windows 11 Pro with 64GB RAM. The processing speed of this new setup turned out to be slower, so the timeout time for each instance was increased to 3600 seconds (up from 1000 sec). We ran the experiments for each MQPF algorithm and the SMT-CBS algorithm, recording data such as the SWAP depths, SWAP gate counts, runtimes and fidelities for each hardware layout. The results for the 8x8 Grid, \textit{ibmq\_rochester} and \textit{ibmq\_paris} layouts are plotted in Figure~\ref{fig:comparison-summary}. Among completed instances, the SWAP depths of solutions found using MQPF algorithms are always lower than or equal to that of the SMT-CBS algorithm. This is expected since the SMT-CBS algorithm optimises the sum of timesteps for each abstract qubit to reach their destination, not the depth. The results show that the optimal and near-optimal MQPF algorithms perform well with respect to the SWAP gate counts, with numbers similar and often lower than solutions obtained from the SMT-CBS algorithm. This is likely because MQPF optimises an objective function which significantly weights each SWAP gate through it's error contribution, thus incentivising solutions with lower numbers of SWAP gates. However, SWAP-gate minimisation is still limited by the algorithm prioritising optimal SWAP depths. As for algorithm runtimes, the log-scale gradient with respect to the number of abstract qubits is significantly smaller for the MQPF algorithm than for the SMT-CBS algorithm. At small instances below 9-12 abstract qubits (20-23 for the 8x8 Grid), the SMT-CBS algorithm has faster runtimes by up to roughly 3 orders of magnitude ({\raise.17ex\hbox{$\scriptstyle\sim$}}0.1 ms compared to {\raise.17ex\hbox{$\scriptstyle\sim$}}10 ms). However, for larger instances with more than 9-12 abstract qubits, the roles are reversed with the MQPF algorithm being faster by up to roughly 3 orders of magnitude ({\raise.17ex\hbox{$\scriptstyle\sim$}}1 sec compared to {\raise.17ex\hbox{$\scriptstyle\sim$}}3600 sec in some cases). This large difference in runtime scaling is likely due to the number of collisions occurring at different densities of abstract qubits, which significantly affects the runtime of the SMT-CBS algorithm. This would also explain why the SMT-CBS algorithm runtime is relatively faster on the 8x8 grid than the other layouts, because random instances on the larger layout with more edges are less likely to have abstract qubits collide. The fidelity data are calculated as $1-(\text{total accumulated error})$ which includes both idle and SWAP gate sources of noise. The optimal and near-optimal MQPF algorithms both find solutions with significantly higher fidelities than the solutions found using the SMT-CBS and MQPF feasible algorithms. This is unsurprising since SMT-CBS and MQPF feasible do not consider optimisation of the objective function and stop when any valid solution is found. The SMT-CBS algorithm finds solutions with marginally higher fidelities on average than the MQPF feasible algorithm, suggesting that only minimising the sum of abstract qubit path lengths (which include idle time steps) is more effective on average than only minimising the SWAP-gate depth. It is interesting to note that on the larger 8x8 Grid and \textit{ibmq\_rochester} layouts, the near-optimal MQPF algorithm is significantly faster than the optimal, however the fidelities of solutions remain close, suggesting that near-optimal MQPF could be a more reliable general-use algorithm with respect to runtime than optimal MQPF. The supplementary material includes additional figures showing comparison data for the other hardware layouts as well as comparative visualised solutions of four problem instances obtained from the optimal MQPF and SMT-CBS algorithms.

By including both idle and swap movement sources of noise, we can investigate the relative contribution of idle errors towards the total accumulated error for each algorithm. In Table~\ref{table:idle-errors}, we show the average ratios of accumulated idle error $\mathcal{E}_\text{idle}$ compared to accumulated SWAP gate error $\mathcal{E}_\text{SWAP}$ among instances with the specified abstract qubit count for each device, calculated as $\mathcal{E}_\text{idle}/\mathcal{E}_\text{SWAP}$. This specified qubit count is chosen to be the highest such that all corresponding instances for every algorithm successfully completed before timing out. The ratios depend on a few properties of the solutions, including the SWAP depth, the total number of SWAPs used, and which qubits at each time step are chosen to participate in movement as opposed to remain idle. We can begin to see this relationship when looking at the difference in values for the MQPF variations, which all have the same SWAP depth. Typically, idle error ratios of optimal MQPF solutions are higher than near-optimal MQPF solutions which are higher than feasible MQPF solutions. For fixed SWAP depth, optimal solutions appear to favour prioritising the SWAP gate error contributions rather than idle error contributions, likely due to SWAP gates introducing significantly more error than instead idling the same qubits for that time period. The SMT-CBS algorithm has the highest idle error ratios, which is likely due to the significant amount of idle error introduced by larger SWAP depths. This also explains why the SMT-CBS solution fidelities are only narrowly higher than from feasible MQPF solutions even though the SWAP gate counts are significantly lower.

\setlength{\tabcolsep}{0pt}
\begin{table}[]
\rowcolors{2}{white}{white!75!black!25}
\begin{tabular}{lcccccc}
Device             & \;\;Qubits\;\; & \;\;Abstract Qubits\;\; & \;\;Optimal MQPF\;\; & \;\;Near-optimal MQPF\;\; & \;\;Feasible MQPF\;\; & \;\;SMT-CBS\;\; \\ \hline
Grid 8x8           & 64     & 28              & 0.289        & 0.282             & 0.179         & 0.426   \\
ibmq\_rochester    & 53     & 21              & 0.256        & 0.257             & 0.223         & 0.372   \\
ibmq\_paris        & 27     & 17              & 0.194        & 0.184             & 0.169         & 0.229   \\ 
ibmq\_poughkeepsie\;\;\;\; & 20     & 20              & 0.204        & 0.190             & 0.164         & 0.223   \\
rigetti\_acorn     & 20     & 17              & 0.189        & 0.188             & 0.172         & 0.225   \\
ibmq\_melbourne    & 15     & 15              & 0.157        & 0.151             & 0.124         & 0.163   \\
\end{tabular}
\caption{The average ratio of idle error compared to SWAP gate error for each algorithm among instances at the specified number of abstract qubits. The number of abstract qubits is chosen to be the highest where no instances timed out.\label{table:idle-errors}}
\end{table}

\section{Discussion}\label{sec:mqp-discussion}
In this study, we explored the challenge of enhancing parts of the compilation process for quantum algorithms, focusing particularly on the multi-qubit pathfinding (MQPF) problem. We benchmarked compilation solutions across various hardware layouts and introduced a classical MQPF algorithm that navigates abstract qubits grouped into teams on a quantum hardware layout towards designated target locations. The algorithm aims to minimise the number of time steps required, where each abstract qubit can either remain idle, or swap with an adjacent qubit at each step. As a secondary objective, the algorithm minimises the total accumulated SWAP gate error rate in the resulting MQPF quantum circuit.

The MQPF algorithm, adapted from the binary integer linear program (BILP) formulation of the multi-commodity flow problem, models abstract qubit navigation as a time-expanded graph. Each time step is represented as a layer comprising all nodes in the hardware graph. The edges between these layers denote possible movements between nodes in the hardware graph. The abstract qubits navigate from the initial layer to the final layer by converting the time-expanded graph into a BILP, which is solved using a mathematical programming solver. The number of layers increases until a valid solution is found, determining the optimal SWAP depth for the MQPF instance. The optimisation's objective function corresponds to the accumulated SWAP gate error in the solution quantum circuits.

When applied to various quantum hardware layouts, the algorithm demonstrated that solutions for the 64-qubit grid layout had significantly lower SWAP depths than those for the 53-qubit heavy-hexagon IBM Quantum layout, despite the grid's larger size and higher number of qubits. This is likely due to the grid's higher connectivity, with nodes of degree 3 and 4, compared to the heavy-hexagon graph's nodes of degree 2 and 3. The higher connectivity allows more direct paths to destinations for each qubit. Additionally, the 64-qubit grid exhibited lower accumulated error rates than all other layouts except the 15-qubit \textit{ibmq\_melbourne} device, another grid layout. This observation was initially surprising, considering the expected longer distances between initial and destination locations on larger layouts. However, the increased connectivity likely offers more directional options for abstract qubits to move towards at every location, helping to avoid congestion and circumvent problematic CNOT gates with relatively high error rates. Regarding computational runtime, we found that larger devices are more time-consuming to compute for because the number of variables and constraints is higher. Although, we found that the 64-qubit grid's runtime is similar to that of the 53-qubit heavy-hexagon layout, even though it is larger. This is likely due to lower SWAP depth solutions, resulting in a shallower time-expanded graph and reduced numbers of variables and constraints.

The advantages of the grid layout over the heavy-hexagon layout could advocate for favouring Google grid-like architectures~\cite{boixo2018characterizing, arute2019quantum} over IBM heavy-hexagon architectures. However, Google devices are typically subjected to a crosstalk constraint, which prohibits adjacent qubits from participating in different operations simultaneously~\cite{booth2018comparing}. This constraint could significantly increase optimal SWAP depths, a question worth exploring in future research by modifying the BILP to include a crosstalk constraint, or by limiting the swap movement edges included in the time expanded graph for each time step. Interestingly, IBM Quantum have recently announced a 120-qubit Nighthawk quantum processor which has a grid-like hardware layout~\cite{ibmqroadmap}.

When compared to the state-of-the-art satisfiability modulo theories and conflict-based search (SMT-CBS) algorithm~\cite{surynek2019multi, surynek2019conflict} from the literature that optimises the sum of lengths of abstract qubit paths to their destinations (which include idle time steps), our method's computational runtime appears to scale better on average, achieving up to 3 orders of magnitude reduction for some large problem sizes. Additionally, we find that the optimal MQPF algorithm's solution SWAP gate counts are on par with the SMT-CBS algorithm's, while the fidelities tend to be significantly higher. Leveraging advanced ILP solvers and approximation techniques, our approach provides a possible trade-off between optimality and computation time, as seen in Section~\ref{sec:comparison_with_smtcbs}. Potential further optimisations in the ILP solver could include column generation and Lagrangian relaxation techniques for multi-commodity flow~\cite{weibin2017solving}, given many solution edges exhibit zero flow, indicating room for potential efficiency improvements. As well as fine-tuning the optimiser settings. Additionally, our method offers flexibility in adapting the model to various mapping problems, such as managing higher numbers of destination nodes than source nodes and handling shared target destinations by different teams. 

As a direction for future research, it would be interesting to explore extending this approach to encompass more aspects of general quantum circuit mapping. This could include combining with the qubit allocation problem, where the focus shifts from destination constraints to constraints based on abstract-qubit interactions.

\medskip
\section*{Acknowledgements} \par 
The author would like to thank Sarama Tonetto, Michael Jones, Charles Hill, and Lloyd Hollenberg for valuable discussions and feedback. This work was supported by the University of Melbourne through the establishment of an IBM Quantum Network Hub at the University. This work was supported by an Australian Government Research Training Program Scholarship. The author gratefully acknowledges the support of the University of Melbourne’s Zero Emission Energy Laboratory (ZEE Lab) and the Victorian Higher Education State Investment Fund (VHESIF).

\medskip
\section*{Data Availability Statement}
The code and datasets used in the current study are available at~\url{doi.org/10.5281/zenodo.11379442}.

\medskip

\bibliographystyle{unsrtnat}
\bibliography{references}

\end{document}